\RequirePackage{fix-cm}
\documentclass[twocolumn]{svjour3}          % twocolumn
\smartqed  % flush right qed marks, e.g. at end of proof
\usepackage{caption}
\usepackage{subcaption}
\usepackage{graphicx}
\usepackage[document]{ragged2e}  
\usepackage{hyperref}
\journalname{Soft Computing}
\begin{document}

\title{Benchmarking Multi-Task Learning for Sentiment Analysis and Offensive Language Identification in Under-Resourced Dravidian Languages%\thanks{Grants or other notes
%about the article that should go on the front page should be
%placed here. General acknowledgments should be placed at the end of the article.}
}

\titlerunning{Benchmarking multi-task learning for under-resourced Dravidian languages}     

\author{Adeep Hande \and Siddhanth U Hegde\and  Ruba Priyadharshini\and  Rahul Ponnusamy \and Prasanna Kumar Kumaresan \and Sajeetha Thavareesan\and  Bharathi Raja Chakravarthi 
%etc.
}

\institute{Adeep Hande* \at
             Indian Institute of Information Technology Tiruchirappalli \\
              %Tel.: +91-99008-65787\\ 
              \emph{adeeph18c@iiitt.ac.in}           %  \\
%             \emph{Present address:} of F. Author  %  if needed
            \and
          Siddhanth U Hegde \at
              University Visvesvaraya College of Engineering, Bangalore University\\ 
              \emph{siddhanthhegde227@gmail.com}
            \and
            Ruba Priyadharshini \at 
            ULTRA Arts and Science College, Madurai Kamaraj University, Madurai, Tamil Nadu, India\\
            \emph{rubapriyadharshini.a@gmail.com}
            \and 
            Rahul Ponnusamy, Prasanna Kumar Kumaresan \at 
            Indian Institute of Information Technology and Management Kerala, India\\
            \emph{\{rahul,prasanna\}.mi20@iiitmk.ac.in}
            %Prasanna Kumar Kumaresan \at 
            %Indian Institute of Information Technology and Management Kerala, India\\
            %\emph{prasanna.mi20@iiitmk.ac.in}
            \and 
            Sajeetha Thavareesan \at
            Eastern University, Sri Lanka\\
            \emph{sajeethas@esn.ac.lk} 
            \and
            Bharathi Raja Chakravarthi* \at
            Insight SFI Research Centre for Data Analytics, National University of Ireland Galway\\
            \emph{bharathi.raja@insight-centre.org}
}

\date{Received: date / Accepted: date}
% The correct dates will be entered by the editor

\maketitle
\justifying 
\begin{abstract}
 
To obtain extensive annotated data for under-resourced languages is challenging, so in this research, we have investigated whether it is beneficial to train models using multi-task learning. Sentiment analysis and offensive language identification share similar discourse properties. The selection of these tasks is motivated by the lack of large labelled data for user-generated code-mixed datasets. This paper works on code-mixed YouTube comments for Tamil, Malayalam, and Kannada languages. Our framework is applicable to other sequence classification problems irrespective of the size of the datasets. Experiments show that our multi-task learning model can achieve high results compared with single-task learning while reducing the time and space constraints required to train the models on individual tasks. Analysis of fine-tuned models indicates the preference of multi-task learning over single-task learning resulting in a higher weighted F1-score on all three languages. We apply two multi-task learning approaches to three Dravidian languages: Kannada, Malayalam, and Tamil. Maximum scores on Kannada and Malayalam were achieved by mBERT subjected to cross-entropy loss and with an approach of hard parameter sharing. Best scores on Tamil was achieved by DistilBERT subjected to cross-entropy loss with soft parameter sharing as the architecture type. For the tasks of sentiment analysis and offensive language identification, the best-performing model scored a weighted F1-score of (66.8\% and 90.5\%), (59\% and 70\%), and (62.1\% and 75.3\%) for Kannada, Malayalam, and Tamil on sentiment analysis and offensive language identification, respectively.  The data and approaches discussed in this paper are published in Github\footnote{\href{https://github.com/SiddhanthHegde/Dravidian-MTL-Benchmarking}{Dravidian-MTL-Benchmarking}}.

%\PACS{PACS code1 \and PACS code2 \and more}
% \subclass{MSC code1 \and MSC code2 \and more}
\end{abstract}

\section{Introduction}

Nowadays, the internet is key to accessing information and communicating with others. The easy access to the internet has enabled the majority of the population to use social media and amplify their communication and connections throughout the world.
Social media platforms such as Facebook, YouTube, and Twitter have paved the way to express any sentiment about the content posted by its users % in social media platform right away in any language
in any language\cite{severyn-etal-2014-opinion,clarke-grieve-2017-dimensions,tian-etal-2017-facebook}.
These texts are informal as they are written in a spoken tone that does not follow strict grammatical rules. Understanding social media content is lately attracting much attention from the natural language processing (NLP) community \cite{barman-etal-2014-code}, owing to the growing amount of data and active users. Sentiment analysis (SA) refers to the method to extract subjectivity and polarity from text \cite{10.1561/1500000011}. The lack of moderation on social media has resulted in the persistent use of offensive language towards other users on their platforms. Offensive language identification (OLI) is the task of identifying whether a comment contains offensive language or not. Traditionally, the classification of SA or OLI is usually attempted in a mono-lingual and single task. Traditional approaches to this text classification problems are pretty helpful for high resourced languages. However, traditional approaches fail for languages with limited resources, and they also fail on code-mixed text \cite{chakravarthi-etal-2020-corpus}.

Multi-task learning (MTL) is a practical approach to utilise shared characteristics of tasks to improve system performances \cite{Caruana1997}. In MTL, the objective is to utilise learning multiple tasks simultaneously to improve the performance of the system \cite{martinez-alonso-plank-2017-multitask}. Since SA and OLI %shares the properties
are both essentially sequence classification tasks, this inspired us to do MTL. To utilise MTL on Dravidian languages, we have experimented with several recent pretrained transformer-based natural language models on Tamil (ISO 639-3: tam), Malayalam (ISO 639-3: mal), and Kannada (ISO 639-3:kan).

Kannada and Malayalam are among the Dravidian languages predominantly spoken in South India and are also official languages in the states of Karnataka and Kerala, respectively \cite{reddy-sharoff-2011-cross}. The Tamil language has official status in Tamil Nadu of India and countries like Sri Lanka, Singapore, Malaysia, and other parts of the world. Dravidian languages are morphologically rich; along with code-mixing, it becomes even more challenging to process these languages \cite{bhat-2012-morpheme}, and they are under-resourced \cite{prabhu-etal-2020-detection}.

We propose a method to perform MTL to address two main challenges arising when creating a system for user-generated comments from social media. The challenges are:
\begin{enumerate}
    \item \label{code-mixing}\textbf{Code-mixing:} In multi-lingual countries like India, Sri Lanka, and Singapore, the speakers are likely to be polyglottic and often switch between multiple languages. This phenomenon is called code-mixing, and these code-mixed texts are even written in non-native scripts \cite{das-gamback-2014-identifying,bali-etal-2014-borrowing,chakravarthi-etal-2020-sentiment}. Code-mixing can be referred to as a blend of two or more languages in a single sentence or conversation, which is prevalent in the social media platforms such as Facebook and YouTube.
    %. Hence, it results in the presence of code-mixed data on social media platforms such as Facebook and YouTube. 

    \item \label{Scarcity of Data} \textbf{Scarcity of Data:}
    Although there is an enormous number of speakers for Tamil, Kannada, and Malayalam, these languages are extensively considered as under-resourced languages. One of the main reasons is the lack of user-generated content extracted from social media applications. One of the ways to tackle the problem is to annotate the data on several tasks.
\end{enumerate}
We address (\ref{code-mixing}) by using pretrained multi-lingual natural language models and (\ref{Scarcity of Data}) the MTL approach.\\
The rest of the paper is organised as follows. Section \ref{Section 2} shows previous work on SA, OLI , and MTL in NLP. Section \ref{Section 4} consists of a detailed description of the datasets for our purposes. Section \ref{section 5} talks about the proposed models for single-task models and their loss functions. Popular MTL frameworks are described in section \ref{Section 6}. We describe the experimental results and analysis in Section \ref{Section 7} and conclude our work in Section \ref{Section 8}.

\section{Related Work}
\label{Section 2}
In this section, we briefly review previous relevant work related to (i) SA, (ii) OLI, and finally, (iii) MTL in NLP.

\subsection{{Sentiment Analysis}}
SA is one of the leading research domains devoted to analyse people's sentiments and opinions on any given entity. Due to its broader applications, there has been a plethora of research performed in several languages. However, the same is not true for the Dravidian languages. As stated earlier, there is a significant lack of data to conduct experiments on code-mixed data in the Dravidian languages. SA is one of the most prominent downstream tasks of NLP as it is essential to obtain people's opinion, which has several business applications in the e-commerce market.

A data set was created as a part of a shared task on SA of Indian languages (SAIL), which consisted of around 1,663 code-mixed tweets extracted from Twitter in Tamil \cite{10.1007/978-3-319-26832-3_61}, where SentiWordNet outperformed all of the other systems \cite{phani-etal-2016-sentiment,das-bandyopadhyay-2010-sentiwordnet}.
There has been a plethora of research performed on several downstream tasks in other languages, primarily due to the abundance of user-generated data in social media, which has developed an interest in people's opinions and emotions with respect to a specific target. Existing research is relatively low due to the lack of data in code-mixing. SA is one of the downstream tasks that are performed on any natural language model. The largely available crowd-sourced data on social media applications such as Twitter and YouTube have resulted in developing several code-mixed datasets for the task.

To our knowledge, a very few Kannada-English code-mixed datasets exist on SA. A Kannada-English code-mixed data set for the emotion prediction was created \cite{appidi-etal-2020-creation}. A probabilistic approach was employed to classify parts of speech (POS) tags \cite{8447784}. Several research pursuits have been worked upon SA in Tamil \cite{se2016predicting,senti}. A recursive neural network approach was opted to improve the accuracy of texts in Tamil \cite{8089122}. A dynamic mode decomposition (DMD) method with random mapping was developed for SA on the SAIL 2015 data set \cite{kumar2020dynamic}. A Lexicon-based approach was employed along with several feature representation approaches to analyse the sentiments on Tamil texts \cite{9063341}. When it comes to Malayalam, several supervised machine learning and rule-based approaches were used to analyse the sentiments of the Malayalam movie reviews \cite{nair2015sentiment,6968548,soumya2020sentiment}. A fuzzy logic-based hybrid approach was also used to analyse the movie reviews in Malayalam \cite{anagha2015fuzzy}.

\subsection{Offensive Language Identification}
A surge in the popularity of social media platforms has resulted in the rise of trolling, aggression, hostile, and abusive language, which is a concerning issue pertaining to the positive/negative impacts a message can imply on an individual or groups of people \cite{tontodimamma2021thirty,PLAZADELARCO2021114120}. This issue has led several researchers to work on identifying offensive language/posts from social media to moderate content on social media platforms to promote positivity \cite{chakravarthi-2020-hopeedi}. Offensive language can be defined as any text entailing certain forms of unacceptable language, which may include insults, threats, or bad words \cite{PLAZADELARCO2021114120}. In comparison, hate speech seems indistinguishable to offensive language. The former aims to detect `abusive' words, which are considered a type of degradation \cite{nobata2016abusive,djuric2015hate}.

There are several ways to detect offensive language. A supervised learning technique was used \cite{10.1007/978-3-642-36973-5_62}, which was based on three decisive factors: content, cyberbullying, and user-based features to tackle cyberbullying. A multi-level classification system was developed that extracts features at different conceptual levels and applies pattern recognition to detect flames (rants, taunts, and squalid phases) in sentences \cite{razavi2010offensive}. It can also be detected by ferreting out offensive and toxic spans in the texts. A toxic span detecting system was developed by leveraging token classification and span prediction techniques that are based on bidirectional encoder representations from transformers {(BERT)} \cite{chhablani2021nlrg}. Multi-lingual detection of offensive spans {(MUDES)} \cite{ranasinghe2021mudes} was developed to detect offensive spans in texts. Several systems were developed to identify offensive language as a part of shared tasks conducted to stimulate research in this domain for Arabic, Danish, English, Greek, and Turkish \cite{zampieri-etal-2019-semeval,zampieri-etal-2020-semeval}. Consequentially, several NLP researchers have worked on developing systems to detect hate speech on social media \cite{trac-2020-trolling,zampieri-etal-2019-semeval}. However, most of the work done on OLI is language-specific, focusing on mono-lingual users in lieu of multi-lingual users, which entails code-mixed text on its social media users \cite{bali-etal-2014-borrowing}. 

For code-mixed sentences, certain researchers analysed the existing state-of-the-art (SoTA) hate speech detection on Hindi-English \cite{rani2020comparative}, while other researchers compared the existing pretrained embeddings for convolutional neural networks {(CNN)} \cite{banerjee2020comparison}. When it comes to OLI, several systems were developed as a part of shared task for OLI in Dravidian languages \cite{dravidianoffensive-eacl,nikhiloffen,adeepoffensive} and Indo-European languages \cite{10.1145/3441501.3441517}. 

\subsection{{Multi-Task Learning}}
% Dude don't start multi-task with sentiment analysis. what do you say?
The primary objective of the MTL model is to improve the learning of a model for a given task by utilising the knowledge encompassed in other tasks, where all or a subset of tasks are related \cite{zhang2018survey}. The essence of MTL is that solving many tasks together provides a shared inductive bias that leads to more robust and generalisable system \cite{changpinyo-etal-2018-multi}. It has long been studied in the domain of machine learning. It has applications on neural networks in the NLP domain \cite{Caruana1997}. However, to the best of our knowledge, MTL models have not been developed for the Dravidian languages yet.
%But no MTL are models developed for Dravidian languages yet.

%Areas in computer vision have a variety and unique model designs 
There is a considerable variety and unique model designs in the field of computer vision since the requirements in the domain of computer vision are vast. Tasks such as image classification, object localisation, object segmentation \cite{8408520}, and object tracking are widespread. Moreover, multi-tasking in videos for real-time applications \cite{8677269} play a vital role in day-to-day life. Multi-task models have also been used for medical images \cite{9210054}.

%The classification task during SA declines when the input contains review comments for multiple tasks. It becomes difficult for the single-task learning (STL) models to perform extremely better on such tasks as they have to work individually. Different techniques have been evolved in recent times. 
MTL models are usually designed with shared encoders that can be customised for the preferred tasks. A multi-scale approach was used to combine long short-term memory {(LSTM)} and CNN for better results as it adds up the benefits of both CNN (nearby sentiments) and LSTM (sentiments that are further away) \cite{9076160}. Attention-based mechanisms can also be used in the encoder, such as multi-layer bidirectional transformer encoder and knowledge encoder that injects knowledge into the language expressions \cite{9136689}. Not only LSTM but also gated recurrent units (GRUs) can be used as a part of multi-task model based on specific tasks; although GRUs have much simpler architecture than the LSTMs, they have the capabilities to forecast futuristic values based on the previous values \cite{9080108}. Such model preferences can be opted to the domain of NLP in order to make the model lighter and faster. The most recent transformation includes using multi-tasks over semi-supervised learning \cite{9226421}, by stacking recurrent neural networks and utilising a sliding window algorithm the sentiments are transferred on to the next item. An empirical study on MTL models for varieties of biomedical and clinical NLP tasks on BERT was proposed \cite{peng-etal-2020-empirical}.

MTL models for the domain of NLP are less in number. The ideas that were applied to other domains such as computer vision, time series, and semi-supervised learning can be acquired, and models could be improved in NLP. As such models are rarely seen, this paper introduces MTL for SA and OLI incorporated from other domains of deep learning.

\section{Dataset}
\label{Section 4}
We make use of the multi-lingual dataset, DravidianCodeMix\footnote{\url{https://github.com/bharathichezhiyan/DravidianCodeMix-Dataset} }\cite{chakravarthi-etal-2021-lre}, consisting of over 60,000 manually annotated YouTube comments. The data set comprises sentences from 3 code-mixed Dravidian languages: Kannada \cite{hande-etal-2020-kancmd}, Malayalam \cite{chakravarthi-etal-2020-sentiment}, and Tamil \cite{chakravarthi-etal-2020-corpus,dravidianoffensive-eacl}.
Each comment is annotated for the tasks of SA and OLI. The code-mixed Kannada dataset consists of 7,273 comments, while the corresponding Malayalam and Tamil code-mixed datasets consist of 12,711 and 43,349 comments, respectively. After removing repetitive sentences, the class-wise distribution of the datasets are specified in Table\,\ref{tab:dataset_distributions} that are to be split into train, validation, and test sets. 

\begin{table*}[htbp]
    \centering 
     
    \begin{tabular}{|l|l|r|l|r|}
    \hline
    \multicolumn{5}{ |c|}{\textbf{Kannada}}  \\
    \hline
     \multicolumn{3}{ |c|}{\textbf{Sentiment analysis}} & \multicolumn{2}{ |c|}{\textbf{Offensive language identification}}
    \\ \hline
    Sl. No. & Class & Distribution & Class & Distribution\\
    \hline
    1 & Positive & 3,291 & Not offensive & 4,121 \\
    2 & Negative & 1,481 & Offensive untargeted & 274 \\
    3 & Mixed feelings & 678 & Offensive targeted individual & 624\\
    4 & Neutral & 820 & Offensive targeted group & 411 \\
    5 & Other language & 1,003 & Offensive targeted others& 145 \\
    6 &- & - & Other anguages & 1,698\\ 
    \hline
    & Total & 7,273 &Total & 7,273 \\
    \hline
        \multicolumn{5}{ |c|}{\textbf{Tamil}}  \\
    \hline
     \multicolumn{3}{ |c|}{\textbf{Sentiment analysis}} & \multicolumn{2}{ |c|}{\textbf{Offensive language identification}}
    \\ \hline
    Sl. No. & Class & Distribution & Class & Distribution\\
    \hline
    1 & Positive & 24,501 & Not offensive & 31,366 \\
    2 & Negative & 5,190 & Offensive untargeted & 3,594 \\
    3 & Mixed feelings & 4,852 & Offensive targeted individual & 2,928\\
    4 & Neutral & 6,748 & Offensive targeted group & 3,110 \\
    5 & Other languages & 2,058 & Offensive targeted others & 582  \\
    6 & - & - & Other languages & 1,769\\ 
    \hline
    & Total & 43,349 &Total & 43,349 \\
    \hline
        \multicolumn{5}{ |c|}{\textbf{Malayalam}}  \\
    \hline
     \multicolumn{3}{ |c|}{\textbf{Sentiment analysis}} & \multicolumn{2}{ |c|}{\textbf{Offensive language identification}}
    \\ \hline
    Sl. No. & Class & Distribution & Class & Distribution\\
    \hline
    1 & Positive & 5,565 & Not offensive & 11,357 \\
    2 & Negative & 1,394 & Offensive untargeted & 171 \\
    3 & Mixed feelings & 794 & Offensive targeted individual & 179\\
    4 & Neutral & 4,063 & Offensive targeted group & 113 \\
    5 & Other languages & 955 &  Other languages & 951\\ 
    \hline
    & Total & 12,771 &Total & 12,771\\
    \hline
    \end{tabular}
     \caption{Class-wise distribution of the datasets for Kannada, Malayalam, and Tamil}
    \label{tab:dataset_distributions}
\end{table*}

The class labels in the dataset are as follows:\\
%\begin{figure*}[ht]
%\centering\includegraphics[width=\textwidth,height=9cm]{doc/example-data.PNG}
%\caption{Example of code-mixing in our Kannada data set}
%\end{figure*} 
\textbf{Sentiment analysis:}
\begin{itemize}
    \item \textbf{Positive state:} Comment contains an explicit or implicit clue in the text suggesting that the speaker is in a positive state.
    \item \textbf{Negative state:} Comment contains an explicit or implicit clue in the text suggesting that the speaker is in a negative state. 
    \item \textbf{Mixed feelings:} Comment contains an explicit or implicit clue in both positive and negative feeling.
    \item \textbf{Neutral state:} Comment does not contain an explicit or implicit indicator of the speaker’s emotional state.
    \item \textbf{Not in intended language:} For Kannada, if the sentence does not contain Kannada written in Kannada script or Latin script, then it is not Kannada.
\end{itemize}
\textbf{Offensive language identification}:
\begin{itemize}
    \item  \textbf{Not offensive}: Comment does not contain offence or profanity.
    \item  \textbf{Offensive untargeted }: Comment contains offence or profanity without any target. These are comments that contain unacceptable language that does not target anyone.
    \item  \textbf{Offensive targeted individual}: Comment contains offence or profanity that targets the individual.
    \item  \textbf{Offensive targeted group}: Comment contains offence or profanity that targets the group.
    \item  \textbf{Offensive targeted other}: Comment contains offence or profanity that does not belong to any of the previous two categories (e.g., a situation, an issue, an organisation, or an event). 
    \item  \textbf{Not in indented language}: Comment is not in the Kannada language.
\end{itemize}
The overall class types are similar in all languages. 
The code-mixed datasets of Kannada and Tamil consist of six classes in OLI, while Malayalam consists of five classes. There is an absence of offensive language others (OTO) class in the Malayalam dataset.

\section{{Methodology}}
\label{section 5}
We explore the suitability of several NLP models on the task of sequence classification. Several pretrained multi-lingual transformer models are investigated to find the better fit for the code-mixed datasets of Tamil, Kannada, and Malayalam. For our purpose, we define single-task learning (STL) models when we train the language models on both of the tasks separately. In this section, we discuss several pretrained transformer-based models that have been used for both STL and MTL. We have implemented the proposed STL and MTL models\footnote{\url{https://github.com/SiddhanthHegde/Dravidian-MTL-Benchmarking}} with the Pytorch library.
%So, we have to prove that method works for Kannada with empirical results.

\subsection{{Transformer-Based Models}}
Recurrent models such as LSTMs and GRUs fail to achieve SoTA results due to longer sequence lengths, owing to memory limitations while batching. While factorisation and conditional computational approaches \cite{shazeer2017outrageously,DBLP:journals/corr/KuchaievG17} have improved the efficiency of the model, however, the underlying issue of computing it sequentially persists.
To overcome this, a transformer was proposed \cite{vaswani2017attention}, an architecture that completely shuns recurrence and restores to attention mechanisms. It is found that adapting to an architecture with attention mechanisms proves to be much more efficient than recurrent architectures. The transformer block follows a stacked encoder-decoder architecture with multi-headed attention and feed forward layers. Self-attention is an attention mechanism relating distinct arrangements of a single sequence to compute a representation of a given sequence.

Scaled dot-product attention is mathematically computed using three vectors from each of the encoder's input vectors: \emph{Query, Key}, and \emph{Value} vectors. \emph{Key} and \emph{Value} assume dimensions \(d_{k}\) and \(d_{v}\), respectively. A softmax function is applied on the dot product of queries and keys in order to compute the weights of the values. Practically, the attention function is computed simultaneously on a set of queries and then stacked into a matrix \emph{Q}. In practice, the attention function is computed on a set of queries simultaneously, being packed into a matrix \emph{Q}. The \emph{Keys} and \emph{Values} are packed into matrices \emph{K} and \emph{V}. The matrix of outputs is computed as follows:
\begin{equation}
\Centering Attention(Q,K,V) = softmax(\frac{QK^T}{\sqrt{d_k}})V
\end{equation}

The above dot-product attention is preferred over additive attention owing to its practical efficiency in both space and time complexities. Self-attention is computed several times in transformer's architecture; thus, it is referred to as multi-head attention. This approach collectively attends information from different representations at different positions. 

\begin{table*}[ht]
 
\begin{tabular}{crrrcrrr}
\noalign{\smallskip}\hline
Sentiment analysis & Kannada & Malayalam & Tamil & Offensive language identification & Kannada & Malayalam & Tamil\\
\noalign{\smallskip}\hline
Positive & 334 & 544 & 2,426 & Not offensive & 407 & 1,134& 3,148 \\
Negative & 164 & 135 & 529 & Offensive untargeted & 27 & 28 & 336 \\
Mixed feelings & 63 & 81 & 465& Offensive targeted individual & 82 & 16 & 316 \\
Neutral & 80 & 424 & 702 & Offensive targeted group & 44 & 10 & 305 \\
Other language & 87 & 94 & 213 & Offensive targeted others& 14 & – & 52 \\
– & – & – & – & Other language & 154 & 90 & 178 \\
\noalign{\smallskip}\hline
Total & 728 & 1,278 & 4,335& & 728 & 1,278 & 4,335 \\
\noalign{\smallskip}\hline
\end{tabular}
\caption{Class-wise distribution of the test sets of Kannada, Malayalam, and Tamil datasets}\label{tbl1}
\end{table*}

\begin{figure}[ht]
\centering\includegraphics[width=6cm,height=12cm]{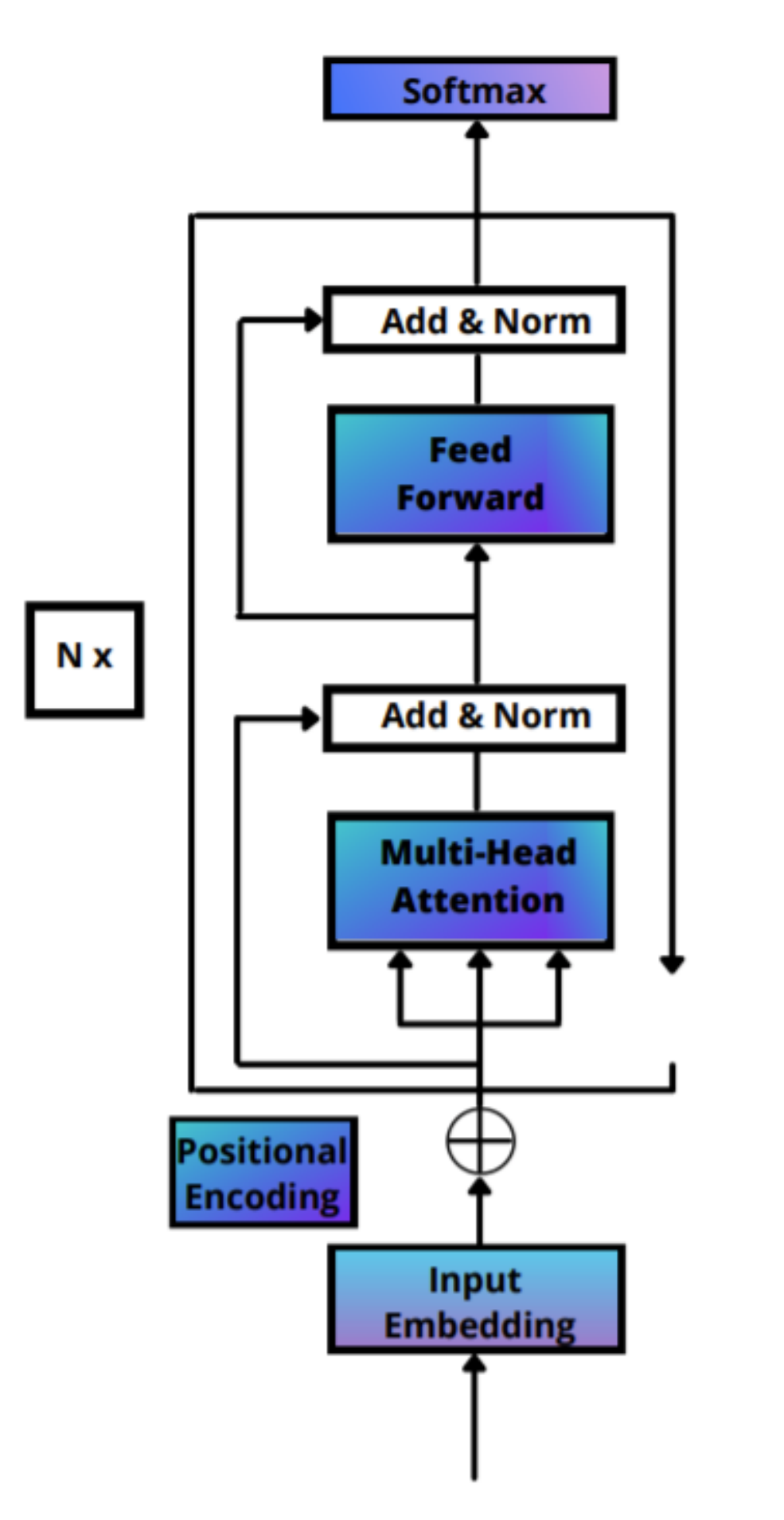}
\caption{ A transformer architecture by \cite{vaswani2017attention} }
\label{Fig: Transformer block}
\end{figure}

Consider the phrase – \emph{\textbf{Ask Powerful Questions}}. In order to calculate the self-attention of the first word \emph{Ask}, the scores for all words in the phrase with respect to \emph{Ask} is to be computed, which then determines the importance of other words when certain words are being encoded into the input sequence. The scores are divided by 8\((2^3)\), which is the square root of the dimension of the key vector.
The score of the first word is calculated using dot-product attention, with the dot product of the query vector \(q_1\) with keys \(k_1, k_2, and k_3\) of all words in the input sentence. The scores are then normalised using the softmax activation.
The normalised scores are then multiplied by vectors \(v_1, v_2, and v_3\) and summed up to obtain self-attention vector \(z_1\). It is then passed to feed FN as input. The vectors for the other words are calculated in a similar way in dot-product attention.
%\begin{figure}[!ht]
%\centering\includegraphics[width=8cm,height=4cm]{doc/self_att_1.PNG}
%\caption{computing the scores from queries and key vectors}
%\label{Fig: self_att_1}
%\end{figure}
 
%\begin{figure}[!ht]
%\centering\includegraphics[width=8cm,height=4cm]{doc/self_att_2.PNG}
%\caption{computing the softmax activation of scores}
%\label{Fig: self_att_2}
%\end{figure}
The normalised scores are then multiplied by vectors \(v_1, v_2, and v_3\) and summed up to obtain self-attention vector \(z_1\). It is then passed to feed forward network (FFN) as input. The vectors for the other words are calculated in a similar way in dot-product attention.
%\begin{figure}[!ht]
%\centering\includegraphics[width=8cm,height=4cm]{doc/self_att_3.PNG}
%\caption{computing the attention vector \(Z_1\)}
%\label{Fig: self_att_3}
%\end{figure}
%\end{itemize}
Softmax is an activation function that transforms the vector of numbers into a vector of probabilities. We use softmax function when we want a discrete variable representation of the probability distribution over \emph{n} possible values. It is a generalisation of sigmoid activation function that is used to represent the probability distribution over a binary variable. Softmax activation function over \emph{K} classes is represented as follows:
\begin{equation}
 softmax (z) = \frac{e^{z_i}}{\sum_{j=1}^K e^{z_j}} 
\end{equation}Whereas, sigmoid is a binary representation of softmax activation function. It is given by:
\begin{equation}sigmoid(x) = \frac{e^x}{e^x + 1}\end{equation}

Along with attention sub-layers in the transformer block, a fully connected FFN persists in each of the layers in its encoder and decoder. It consists of a rectified linear unit {(ReLU)} activation in-between two positions.
\begin{equation}FN(x) = \max(0,xW_1,b_1)W_2 + b_2 \end{equation}
ReLU, a rectified linear activation function, is a piece-wise linear function that will output the input directly if positive, else it will output it as zero. ReLU overcomes the vanishing gradient problem, thus allowing models to learn faster and perform better. It is mathematically computed as follows:
\begin{equation}y = \max (0,x) \end{equation}

While attention mechanism is a major improvement over recurrence-based seq2seq models, it does have its own limitations. As attention can only deal with fixed-length strings, the text has to be split into a number of chunks before feeding them into the inputs. The chunking inadvertently results in context fragmentation. This means that a sentence, if split in the middle, will result in significant loss of context. It would mean that the text is split without considering sentence or any other semantic boundary.

Due to the absence of recurrence or convolutional layers, some information about the tokens' relative or absolute position must be fed for the succession of sequence order in the model. Thus, \emph{Positional Encodings} is added to the input embeddings as shown in Fig. \ref{Fig: Transformer block}. Most of the SoTA language models assume the transformer block as its fundamental building block in each layer.

%\subsection{{STL Models}}
%For our purpose, we define STL Models, as when we train the transformer-based models on both of the tasks separately, the results of which are discussed in the further sections.%as shown in Tables \ref{stl_kan}, \ref{stl_mal}, and \ref{stl_tam}.  

\subsubsection{{BERT}}
\label{BERT}
The computer vision researchers have repeatedly demonstrated the advantages of transfer learning, pretraining a neural network model on a known task  \cite{5206848}. 
%Consider pretraining a model for ImageNet \cite{5206848}, and so fine-tuning, using the pretrained neural network because the basis of a replacement purpose-specific model.% In recent years, researchers are showing that an analogous technique may be useful in many natural language tasks.

BERT, a language representation model is designed to pretrain deep bidirectional representations from unlabelled text by jointly conditioning on both left and right context in all layers \cite{devlin-etal-2019-bert}. It has been trained on 11 NLP tasks. %BERT uses masked language models to enable pretrained deep bidirectional representations. 
%BERT consists of two steps: \emph{pretraining} and \emph{fine-tuning}. During pretraining, the model is trained on unlabelled data over different pretraining tasks.
\begin{figure*}[!ht]
\centering\includegraphics[width=0.9\textwidth,height=8cm]{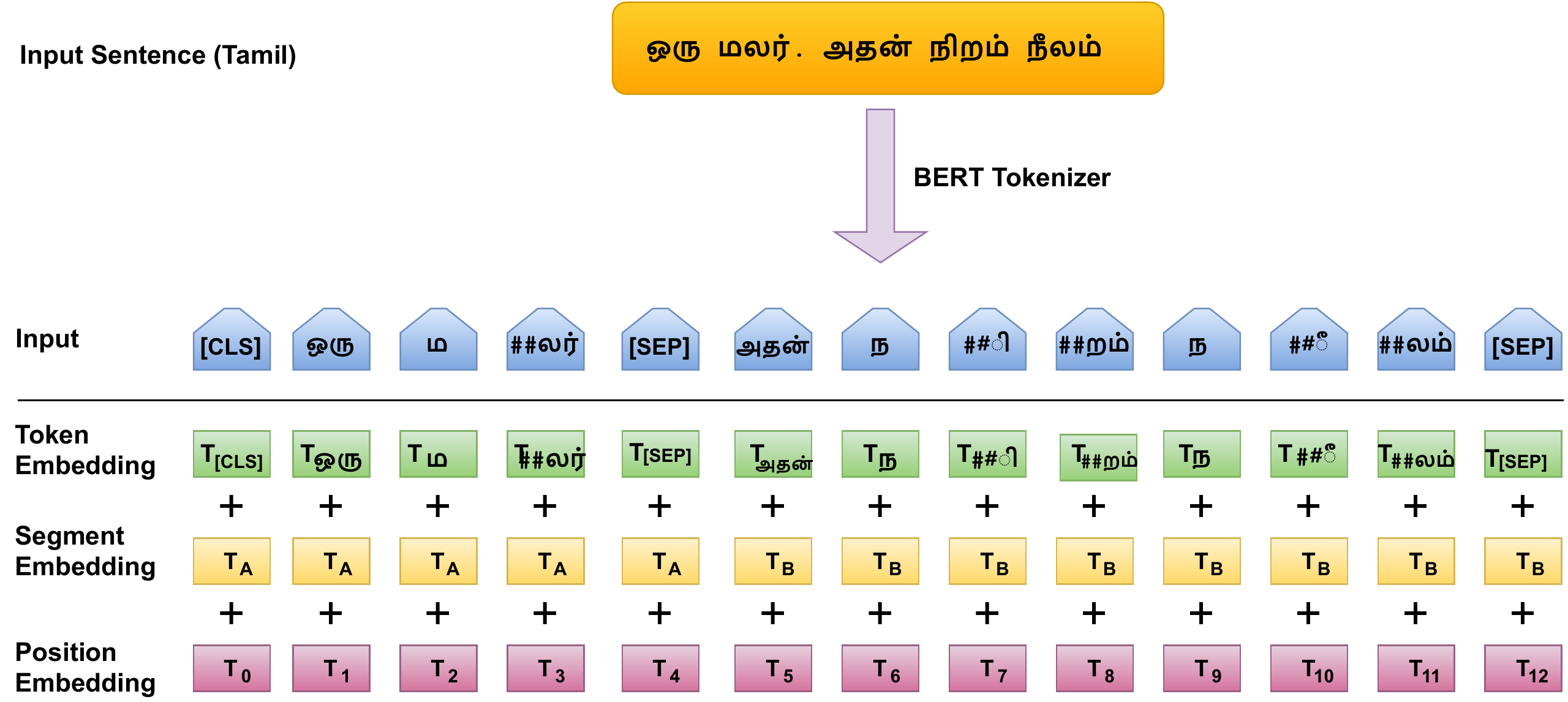}
\caption{ Illustration of BERT input representations. The Tamil sentence translates to {``This is a Flower. It is blue in colour."} The input embeddings are the sum of token, segmentation, and positional embeddings \cite{devlin-etal-2019-bert}}.
\label{Fig: BERT_input_representations}
\end{figure*}
BERT is pretrained on two unsupervised tasks: 
\begin{itemize}

\item {Masked Language Modelling (MLM):}

Standard LMs \cite{Radford2018ImprovingLU} can only be trained left-to-right or right-to-left. However, training it bidirectionally might allow the word to spot itself accidentally. To pretrain deep representations, a certain percentage of the input tokens are masked arbitrarily, then predict those masked tokens, and are referred to as “Masked Language Modeling”, which was previously stated as \emph{Cloze} \cite{Taylor1953ClozePA}. Before feeding word sequences into BERT, 15\% of the words in each sequence are replaced with a [MASK] token. The model then attempts to predict the initial value of the masked words, supported by the context provided by the opposite non-masked words within the sequence.
 
\item {Next-Sentence Prediction (NSP):}

One of the primary drawbacks of LM is its inability to capture the connection between two sentences directly. However, important downstream tasks such as question answering (QA) and natural language interference (NLI) are supported with the relationships between two sentences. To overcome this obstacle pretraining for a \emph{Binarized Next Sentence Prediction} task is performed. Specifically, when choosing the sentences A and B for every pretraining example, 50\% of the time B is that the actual next sentence that follows A (labelled as \emph{Is Next}), and 50\% of the time it is a random sentence from the corpus (labelled as \emph{Not Next}). In Fig. \ref{Fig: BERT_input_representations}, the input consists of two Tamil sentences. After tokenising the sentences, we observe that there are 13 input tokens, including the [CLS] and [SEP] tokens. Tokens \(T_1\), \(T_2\), ..., \(T_13\) represent the positional embeddings while \(T_A\) and \(T_B\) represent the tokens for whether a given sentence follows the other.
\end{itemize}
%\textbf{Loss}:

%In addition to the MLM loss \cite{devlin-etal-2019-bert}, BERT uses an additional loss called NSP. NSP is a binary classification loss to predict if two segments occur consecutively, as follows: positive examples are created by taking consecutive segments from the training corpus; negative examples are created by pairing segments from different documents; and positive and negative examples are sampled with equal probability. The NSP objective was designed to improve performance on downstream tasks, such as natural language inference, that require reasoning about the relationship between sentence pairs.

%For fine-tuning, it is primarily initialised with the pretrained parameters, and all of the parameters are fine-tuned using labelled data from other tasks. Initially, \cite{devlin-etal-2019-bert} reported the results for two model sizes: \(BERT_{BASE}\) (L=12, H=768, A=12, Total Parameters=110M) and \(BERT_{LARGE}\) (L=24, H= 1024, A=24, Total Parameters=340M), where L denotes the number of layers (i.e., transformer blocks); H denotes the hidden size; and A denotes the number of self-attention heads.

\begin{figure}[ht]
\centering\includegraphics[width=10cm,height=8cm,angle=90]{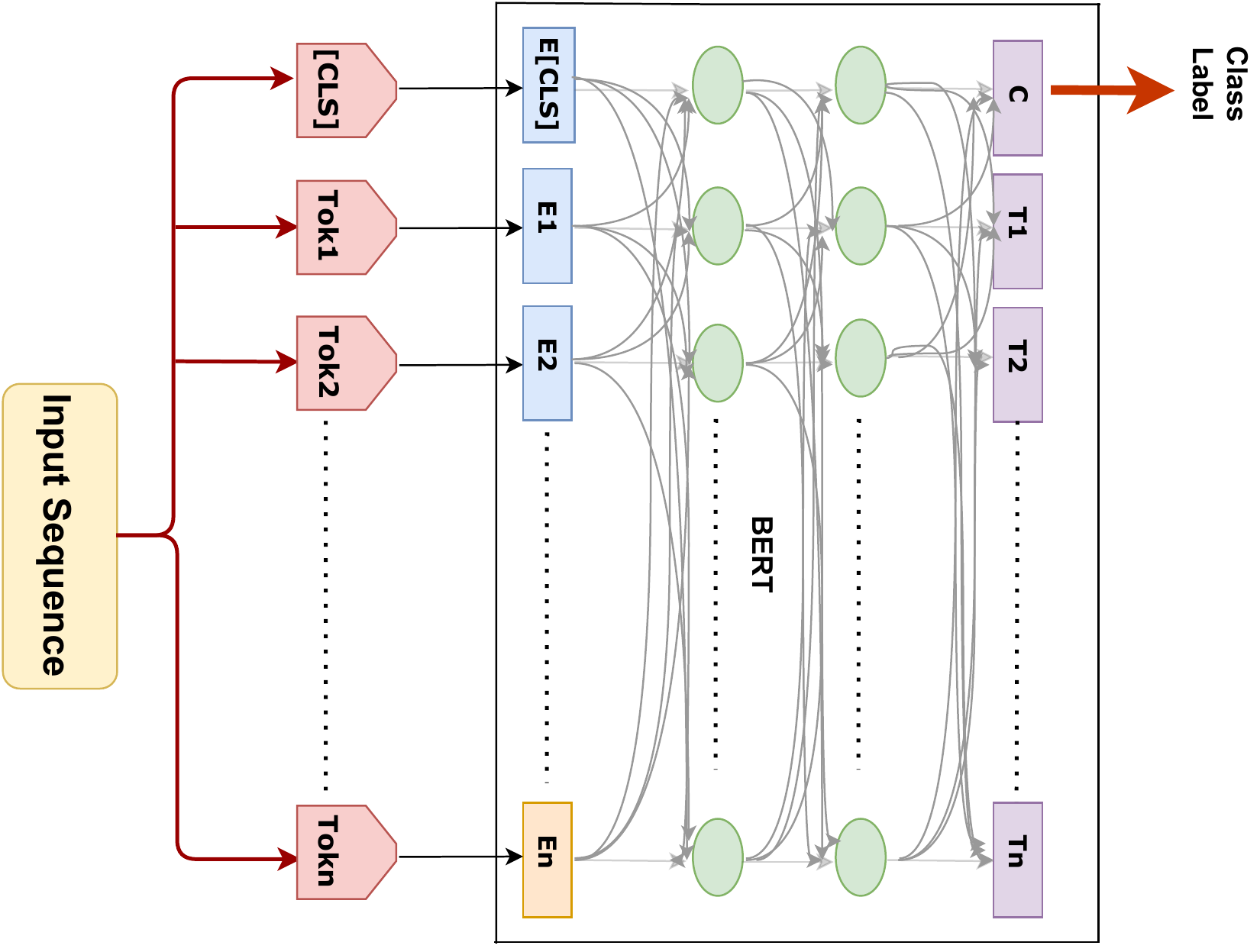}
\caption{Illustration of fine-tuning BERT on single sentence classification tasks such as SA and OLI \cite{devlin-etal-2019-bert}}.
\label{Fig:single_task_pred}
\end{figure}

%For our required NLP task of \emph{Sequence Classification}, we deal with code-mixed data in Kannada. Thus, it would be better to use a BERT model that has been pretrained on Kannada and English. Thus, we will be fine-tuning BERT for this downstream task. 
We use the sequence tagging specific inputs and outputs into BERT and fine-tune all parameters end to end. At the input, sentences A and B from pretraining are equivalent to a degenerate text pair in sequence tagging. The [CLS] representation is fed into an output layer for classification as observed in Fig. \ref{Fig:single_task_pred}.
%We use the pretrained models offered by huggingface transformers \fntext[Huggingface Transformers]{https://huggingface.co/transformers/pretrained_models.html}.% can you check how to add foot-notes
%The primary step is to use the pretrained BERT tokeniser to first cleave the word into tokens. Then, we add the special tokens needed for sentence classifications ([CLS] at the first position and [SEP] at the end of the sentence as shown in Fig. \ref{Fig: BERT_input_representations}).
%After special tokens are added, the tokeniser replaces each token with its ID from the embedding table that is a component we obtain from the pretrained model. The output would be a vector for each input token. Each vector is made up of 768 numbers (as H of \(BERT_{BASE}\) is 768). As this approach is task-specific, the first vector (the one associated with the [CLS] token). This vector is then passed on to a dense layer for classification. 

%We use models that were primarily pretrained on multiple languages. Thus, we use \emph{bert-base-multilingual-cased} and \emph{bert-base-multilingual-uncased} for our tasks. \emph{bert-base-multilingual-cased} (L=12, H=768, A=12, Total Parameters=179M) was trained on cased text in the top 104 languages with the largest Wikipedias. \emph{bert-base-multilingual-uncased} (L=12, H=768, A=12, Total Parameters=168M) was trained on lower-cased text in the top 102 languages with the largest Wikipedias. \fnref{https://github.com/google-research/bert/blob/master/multilingual.md}

\subsubsection{{DistilBERT}}
\label{DistilBERT}
%Due to the increasing adaptation of researchers towards transfer learning from large-scale language representation models, there could be some constraints on the computational resources, which makes fine-tuning challenging. Thus, 
DistilBERT was proposed \cite{DBLP:journals/corr/abs-1910-01108} as a method to pretrain a smaller language representation model that serves the same purpose as other large-scale models like BERT. DistilBERT is 60\% faster and has 40\% fewer parameters than BERT. It also preserves 97\% of BERT performance in several downstream tasks by leveraging the knowledge distillation approach. Knowledge distillation is a refining approach in which a smaller model—the student—is trained to recreate the performance of a larger model—the teacher.
DistilBERT is trained on a triple loss with student initialisation.

Triple loss is a linear combination of the distillation loss \(L_{ce}\) with the supervised training loss, the MLM loss \(L_{mlm}\), and a cosine embedding loss (\(L_{\cos}\)) that coordinates the direction of the student and teacher hidden state vectors, where:

\begin{equation}L_{ce} = \sum_{i}t_{i} * \log(s_{i})\end{equation}
where t\(_i\) (resp. si) is a probability estimated by the teacher (resp. the student) \cite{DBLP:journals/corr/abs-1910-01108}.

We use a pretrained multi-lingual DistilBERT model from the huggingface transformer library \emph{distilbert-base-multilingual-cased} for our purpose, which is distilled from the mBERT model checkpoint.
\subsubsection{{ALBERT}}
\label{ALBERT}
Present SoTA LMs consist of hundreds of millions if not billions of parameters. In order to scale the models, we would be restricted by the memory limitations of compute hardware such as GPUs or TPUs. It is also found that increasing the number of hidden layers in the BERT-large model (340M parameters) can lead to worse performance. Several attempts of parameter reduction techniques to reduce the size of models without affecting their performance. Thus, a lite BERT (ALBERT) for self-supervised learning of language representations was proposed. ALBERT \cite{DBLP:journals/corr/abs-1909-11942} overcomes the large memory consumption by incorporating several memory reduction techniques, factorised embedding parameterisation, cross-layer parameter sharing, and Sentence ordering Objectives.
\begin{itemize}
\item {Factorised embedding parameterisation}

It is learned that WordPiece embeddings are designed to learn context independent representations, but hidden-layer embeddings are designed to learn context dependent representations. BERT heavily relies on learning context dependent representations with the hidden layers. It is found that embedding matrix \emph{E} must scale with hidden layers \emph{H}, and thus, this results in models have billions of parameters. However, these parameters are rarely updated during training indicating that there are insufficient useful parameters. Thus, ALBERT has designed a parameter reduction method to reduce memory consumption by changing the result of the original embedding parameter \emph{P} (the product of the vocabulary size \emph{V} and the hidden layer size \emph{H}). 
\begin{equation} V * H = P \rightarrow{} V * E + E * H = P\end{equation}
where \emph{E} represents the size of the low-dimensional embedding space. In BERT, \emph{E} = \emph{H}. While in ALBERT, {H} \(\gg\) \emph{E}, so the number of parameters will be greatly reduced.
\item\textbf{Cross-layer parameter sharing}

ALBERT aims to elevate parameter efficiency by sharing all parameters, across all layers. Hence, feed FN and attention parameters are shared across all layers as shown in Fig. \ref{Fig: cross_layer}.
\begin{figure}[!ht]
\centering\includegraphics[width=8cm,height=5cm]{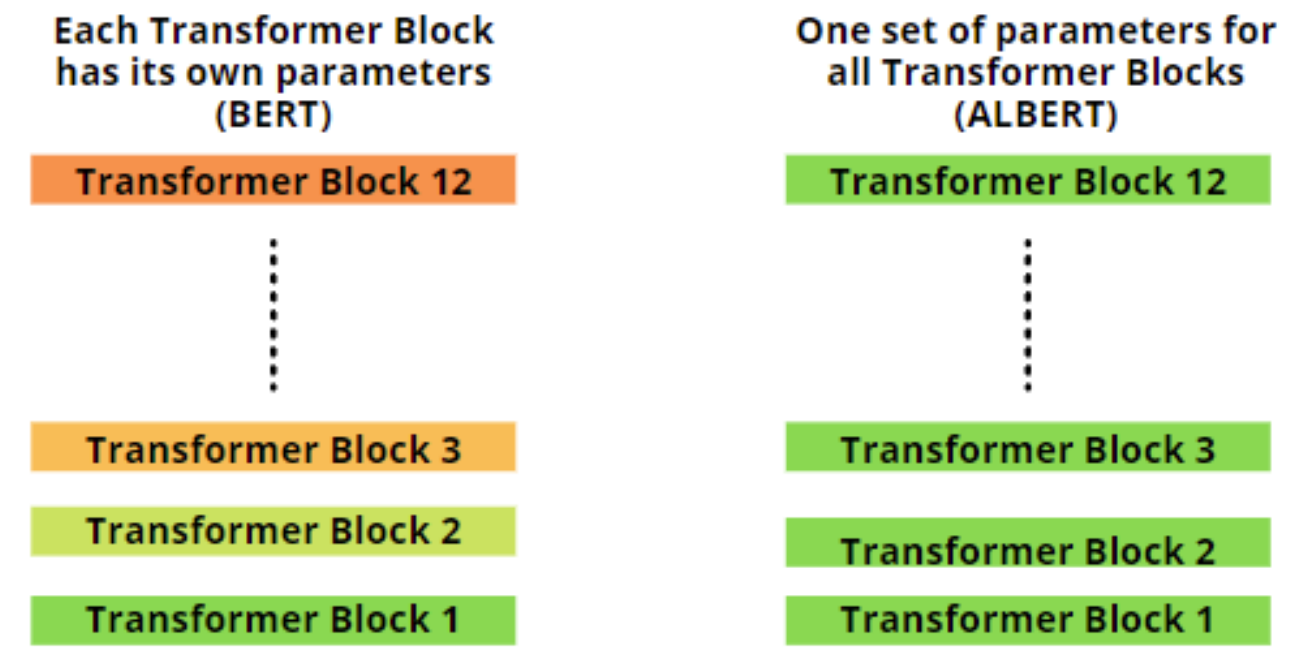}
\caption{ No shared parameters in BERT vs cross-layer parameter sharing in ALBERT}
\label{Fig: cross_layer}
\end{figure}

\item {Sentence order prediction (SOP)}

ALBERT uses MLM, as similar to BERT, using up to 3 word masking (max{(n\_gram) = 3}). ALBERT also uses SOP for computing inter-sentence coherence loss. Consider two sentences being used in the same document. The positive test case is that the sentences are in a correct order, while the negative test case states that the sentences are not in a proper order. SOP results in the model learning finer-grained distinctions about coherence properties, while additionally solving the NSP task to a rational degree. Thus, ALBERT models are more effective in improving downstream tasks' performance for multi-sentence encoding tasks.

\end{itemize}
By incorporating these features and loss functions, ALBERT requires much fewer parameters in contrast to the base and large versions of BERT models proposed earlier in \cite{devlin-etal-2019-bert}, without hurting its performance. 
\subsubsection{{RoBERTa}}
\label{RoBERTa}
 
Robustly optimised BERT (RoBERTa) \cite{DBLP:journals/corr/abs-1907-11692} is a BERT-based model. The difference is within the masking technique. BERT performs masking once during the information processing, which is essentially a static mask; the resulting model tends to see the same form of masks in numerous training phases. RoBERTa was designed to form a dynamic mask within itself, which generated masking pattern changes every time the input sequence is fed in, thus playing a critical role during pretraining. The encoding used was byte-pair encoding (BPE) \cite{sennrich-etal-2016-neural}, which may be a hybrid encoding between character and word-level encoding that allows easy handling of huge text corpora, meaning it relies on subwords instead of full words. The model was made to predict the words using an auxiliary NSP loss. Even BERT was trained on this loss and was observed that without this loss, pretraining would hurt the performance, with significant degradation of results on QNLI, MNLI, and SQuAD.

\subsubsection{{XLM}}
\label{XLM}
The XLM model was proposed in cross-lingual language model pretraining 
\cite{dai-etal-2019-transformer}. This model uses a shared vocabulary for different languages. For tokenising the text corpus, BPE was used. Causal language modelling (CLM) was designed to maximise the probability of a token \(x_t\) to appear at the \emph{t}th position in a given sequence. MLM is when we maximise the probability of a given masked token \(x_t\) to appear at the \emph{t}th position in a given sequence. Both CLM and MLM perform well on mono-lingual data. Therefore, XLM model used a translation language modelling. The sequences are taken from the translation data and randomly masks tokens from the source as well as from the target sentence. Similar to other transformer-based mono-lingual models, XLM was fine-tuned on XLNI data set for obtaining the cross-lingual classification. Downstream tasks on which this was evaluated on were tasks such as cross-lingual classification, neural machine translation, and LMs for low-resource languages. 

\subsubsection{{XLNet}}
\label{XLNet}
BERT performed extremely well on almost every language modelling task. It was a revolutionary model as it could be fine-tuned for any downstream task. But even this came with a few flaws of its own. BERT was built in such a manner that it replaces random words in the sentences with a special [MASK] token and attempts to predict what the original word was. XLNet \cite{DBLP:journals/corr/abs-1906-08237} pointed out certain major issues during this process. The [MASK] token that is used in the training would not appear during fine-tuning the model and for other downstream tasks. However, this approach could further create problems failing to replace [MASK] tokens at the end of pretraining. Moreover, the model finds it difficult to train when there are no [MASK] tokens in the input sentence. BERT also generates the predictions independently, meaning it does not care about the dependencies of its predictions.

XLNet uses autoregressive (AR) language modeling that aims to estimate the probability distribution of a text corpus and without using the [MASK] token and parallel independent predictions. It is achieved through the AR modelling as it provides a reasonable way to express the product rule of factorising the joint probability of the predicted tokens. XLNet uses a particular type of language modelling called the "permutation language modelling" in which the tokens are predicted for a particular sentence in random order rather than sequential order. The model is forced to learn bidirectional model dependencies between all combinations of the input. It is significant to note that it permutes only the factorisation order and not the sequence order, and it is rearranged and brought back to the original form using the positional embedding.
For sequence classification, the model is fine-tuned for sentence classifier, and it does not predict the tokens but predicts the the sentiment according to the embedding. The architecture of the XLNet uses transformer XL as its baseline \cite{DBLP:journals/corr/abs-1901-02860}. The transformer adds recurrence to the segment level instead of the word level. Hence, fine-tuning is carried out by caching the hidden states of the previous states and passing them as keys or values when processing the current sequence. The transformer also uses the notion of relative embedding instead of positional embedding by encoding the relative distance between the words. 

%\begin{figure}
%\centering\includegraphics[width=6cm,height=6cm]{edited xlnet.pdf}
%\caption{: Overview of the
%permutation language modelling training with two-stream attention}
%\label{Fig:XLNet}
%\end{figure}

\subsubsection{{XLM-RoBERTa}}
\label{XLM-RoBERTa}
XLM-RoBERTa was proposed as an unsupervised cross-lingual representation approach \cite{conneau-etal-2020-unsupervised}, and it significantly outperformed multi-lingual BERT on a variety of cross-lingual benchmarks. XLM-R was trained on Wikipedia data of 100 languages and fine-tuned on different downstream tasks for evaluation and inference. The XNLI data set was used for machine translation from English to other languages and vice versa. It was checked on named entity recognition (NER) and cross-lingual QA. It achieved great results on the standart general language understanding evaluation {(GLUE)} benchmark and achieved SoTA results in several tasks.

\subsubsection{{CharacterBERT}}
\label{CharacterBERT}
\begin{figure}[!ht]
\centering\includegraphics[width=8cm,height=10cm]{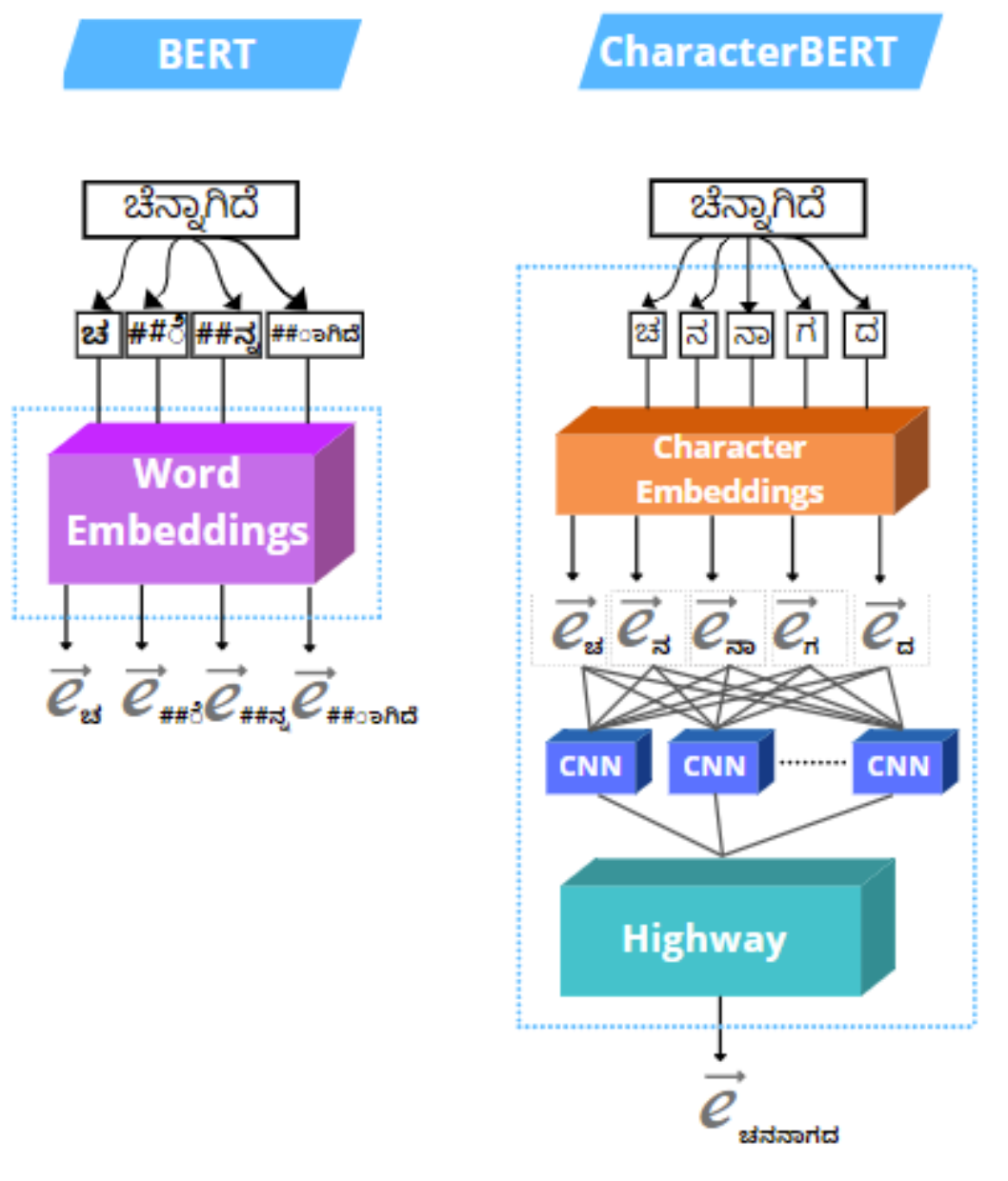}
\caption{Comparison of the context-independent representation systems and tokenisation approach in BERT and characterBERT \cite{el-boukkouri-etal-2020-characterbert}}
\label{Fig:characterBERT}
\end{figure}

The success of BERT has led many language representation models to adapt to the transformers architecture as their primary building block, consequently inheriting the \emph{wordpiece} tokenisation. This could result in an intricate model that focuses on subword rather than word, for specialised domains. Hence, characterBERT, a new variant of BERT that completely drops the \emph{wordpiece} system and uses a Character-CNN module instead of representing the entire words by consulting their characters \cite{el-boukkouri-etal-2020-characterbert} as shown in Fig. \ref{Fig:characterBERT}. 

The characterBERT is based on “base-uncased” version of BERT (L = 12, H = 768, A = 12, and total parameters = 109.5 M). The subsequent characterBERT architecture has 104.6 M parameters. Usage of character-CNN results in a smaller overall model, in spite of using a complex character module, m for BERT’s \emph{wordpiece} matrix possesses 30K X 768-D vectors, while characterBERT utilises 16-D character embeddings with the majority of small-sized CNNs. Seven 1-D CNNs with the following filters are used: [1, 32], [2, 32], [3, 64], [4, 128], [5, 256], [6, 512], and [7, 1024]. The Kannada word {\emph{chennagide}} written in Latin script, which can be translated as \emph{Great}. Fig. \ref{Fig:characterBERT} compares how characterBERT focuses more on the entire word rather than subwords as observed in BERT.

\subsection{{Multi-Task Learning}}
\label{Section 6}
We generally focus more on optimising a particular metric to score well on a certain benchmark. Thus, we train on a single model or an ensemble of models to perform our desired task. These are then fine-tuned until we no longer see any major increase in its performance. However, the downside to this is that we lose out on much of the information that could have helped our model train better. If there are multiple related tasks, we could devise an MTL model that can further elevate its performance in multiple tasks by learning from shared representations between related tasks. We aim to achieve better F1-scores by leveraging this approach. Since both SA and OLI are sequence classification tasks, we believe that an MTL model trained on these two related tasks will achieve better results, in comparison with their performances when trained separately on a model to predict a single task. Most of the methodologies of MTL in NLP are based on the success of MTL models in computer vision. We use two novel approaches for MTL in NLP, which are discussed in the subsequent sections. 
\subsubsection{{Hard Parameter Sharing}}
\begin{figure}[!ht]
\centering\includegraphics[width=8cm,height=7cm]{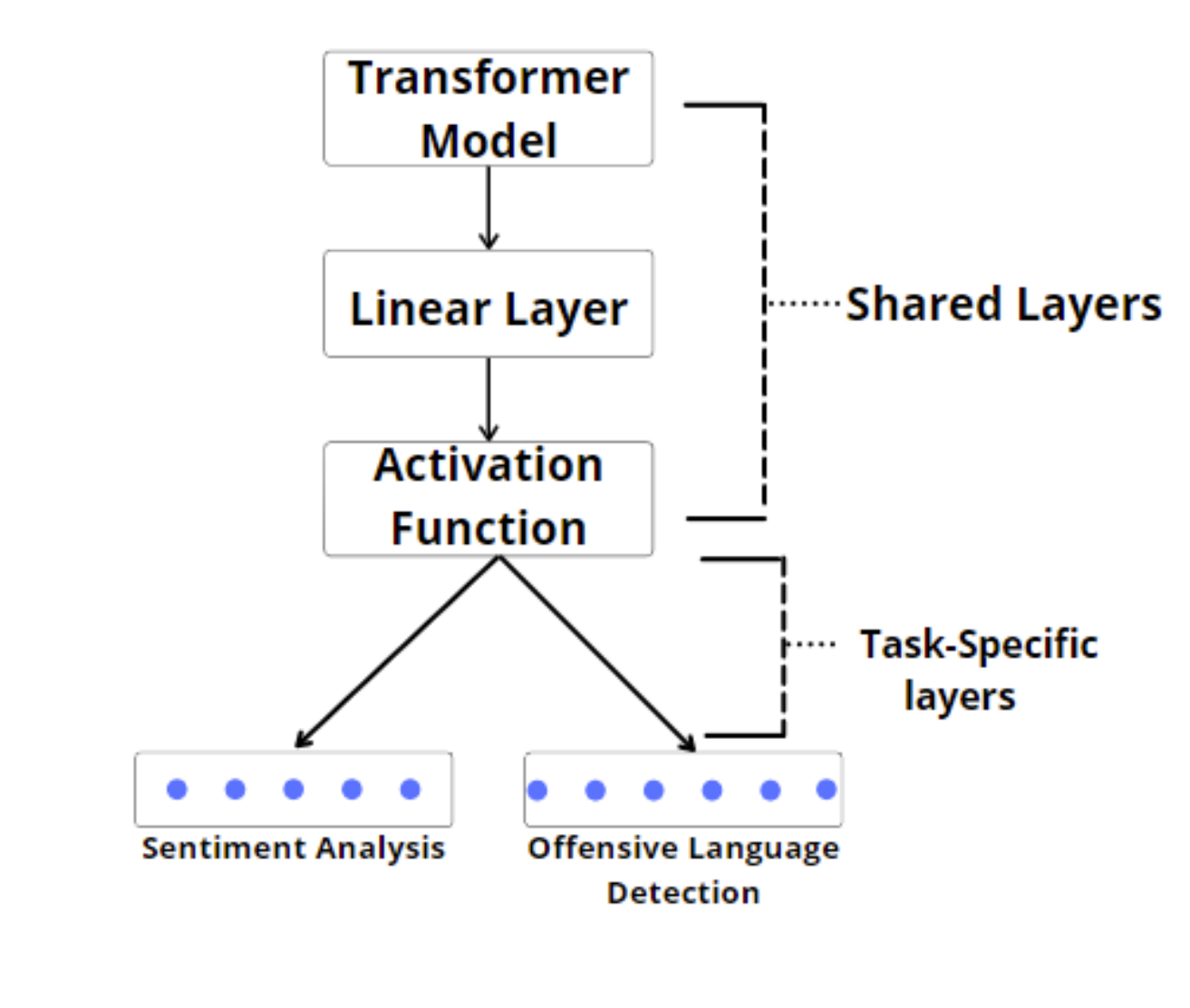}
\caption{Hard parameter sharing for sentiment analysis and offensive language identification }
\label{Fig:Hard Parameter Sharing}
\end{figure}

This is one of the most common approaches to MTL. Hard parameter sharing of hidden layers \cite{Caruana1997} is applied by sharing the hidden layers between the tasks while keeping the task-specific output layers separate. Hard parameter sharing is represented in Fig. \ref{Fig:Hard Parameter Sharing}. As both of the tasks are very similar, one of the tasks of the model is supposed to learn from the representations of the other task. The more the number of tasks that we are learning simultaneously, the more the representations the model will capture and reduce the chances of overfitting in our model \cite{DBLP:journals/corr/Ruder17a}. As shown in Fig. \ref{Fig:Hard Parameter Sharing}, the model will have two outputs. We would be using different loss functions to optimise the performance of the models.
%write about handling imbalance in the class weights
%why adding losses preforms better than individual losses
%why xlm r for mtl over other stl models (cite xlm r if u get)
%cross validation on mtl
%why we didnt do cross validation on stl as it was just to check which model was better

%To train the model, we use two loss functions, both being \emph{nn.CrossEntropyLoss}. This loss function is a softmax activation of the negative loss likelihood function and then calculates the accumulated gradients during training. We calculate the accuracies using three approaches. As both of the tasks consist of multiple classes, we use cross-entropy as the loss function. 

A loss function takes a pair (output and target) of inputs and computes a value that estimates how far away the output is from the target. To back propagate the errors, we first clear the gradients to prevent the accumulation of gradients in each epoch. In this method, we propagate back the losses separately, and the optimiser updates the parameters for each of the tasks separately. After the loss function calculates the losses, we add the losses of the two tasks, and the optimiser then updates the parameters for both tasks simultaneously, where \(Loss_{SA} \) and \( Loss_{OLI}\) are losses for SA and OLI, respectively.
\begin{equation} Loss = Loss_{SA} + Loss_{OLI} \label{naive_addition}\end{equation}  
%The drawback to adding loss functions is that at times usually one of the tasks performs better while suppressing the performance of the other.

\subsubsection{Soft Parameter Sharing}
This is another approach of MTL where constrained layers are added to support resemblance among related parameters. Unlike hard parameter sharing, each task has its own model, and it learns for each task to regularise the distances between the different models' parameters in order to encourage the parameter to be similar. Unlike hard parameter sharing, this approach gives more adaptability to the tasks by loosely coupling the representations of the shared space. We would be using Frobenius norm for our experiments as the additional loss term.%Unlike hard parameter sharing, each task has its own model, the distances between the parameters of the respective models are regularised to have similar parameters 

\begin{figure}[htbp]
\center\includegraphics[width=6cm,height=7cm]{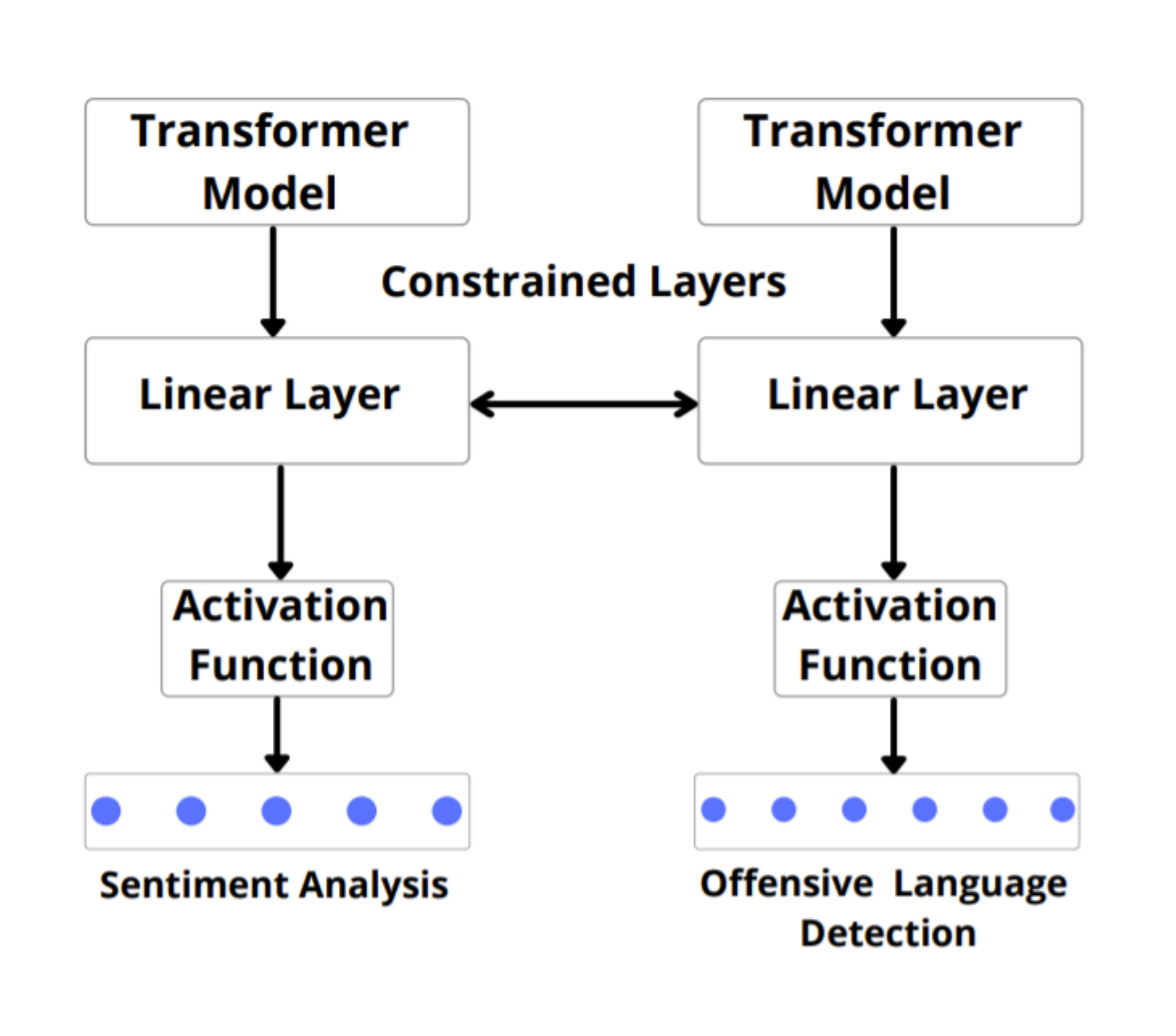}
\center\caption{Soft parameter sharing for sentiment analysis and offensive language identification }
\label{Fig:Soft Parameter Sharing}
\end{figure}

\begin{enumerate}
    \item  {Additional Loss Term}
    
    This approach of soft parameter sharing is to impose similarities between corresponding parameters by augmenting the loss function with an additional loss term, as we have to learn two tasks. Let \(W_i^j\) denote the task j for the \(i^{th}\) layer. 
    \begin{equation}L = l + \sum_{S(L)}\lambda_{i}\mid\mid W_i^{(S)} - W_i^{(O)} \mid\mid_{(F)}^2 \end{equation}
    
    where l is the original loss function for both of the tasks, and \(\mid\mid.\mid\mid_{F}^2\) is the squared Frobenius norm. A similar approach was used for low-resource dependency parsing, by employing a cross-lingual parameter sharing whose learning objective involved regularisation using Frobenius norm \cite{duong-etal-2015-low}.  
    
    \item {Trace Norm}
    
    We intend to penalise trace norm resulting from stacking \(W_i^S\) and \(W_i^O\). A penalty encourages estimation of shrinking coefficient estimates, thus resulting in dimension reduction \cite{Yang2017TraceNR}. The trace norm can be defined as the sum of singular values:
    \begin{equation}\mid\mid\textbf{W}\mid\mid = \sum_i \sigma_i\end{equation}
    
\end{enumerate}
 
\subsection{Loss Functions}
 
Loss function is a technique of evaluating the effectiveness of specific algorithm models on the given data. If the predictions are deviating from expected results, the purpose of a loss function is to output a high value. The output of the loss function decreases when the predictions start to match the expected results. With the help of an optimisation function, a loss function comprehends to reduce the error in the predictions. In this section, we will be discussing several loss functions that are going to be used in our experiments.
\subsubsection{Cross-Entropy Loss}
This is one of the most commonly used loss functions for classification tasks. It quantifies from the study of information theory and entropy and is calculated as the difference of the probability distributions for a given random variable. Entropy can be defined as the number of entities required for the transmission of a randomly selected event in a probability distribution. "Cross-entropy is the average number of bits needed to encode data from a source coming from a distribution \textit{p} when we use a model \textit{q}" \cite{MurphyML}. The divination of this definition can be conveyed if we consider a target as an underlying probability distribution {P} and an approximation of the target probability distribution as {Q}. Let the cross-entropy between two probability distributions, Q from P, be stated as {H(P,Q)}. It is calculated as follows:
\begin{equation}H(P,Q) = \sum_x P(x)\log(Q(x))\end{equation} 
For the sake of multi-label classification, it is computed as follows:
\begin{equation}L_{CE} = \sum_{i}^C t_{i}\log (s_{i}) \end{equation}
where \(t_{i}\) and \(s_{i}\) are the ground truths for each class i in C.\\
\textbf{Considering Class Imbalance}

As specified earlier, we observe that the datasets have class imbalances and are not equally distributed as shown in Table\,\ref{tab:dataset_distributions}. Thus, we consider the weights of each class and pass the weights as a tensor to the parameter while computing the loss. Class Weights give inverse class weights to penalise the underrepresented class for class imbalance.
\begin{equation}
W_i = 1 - \frac{W_i}{\sum_{i}^C W_i} \end{equation}
where C is the number of classes. The resultant tensor is \([w_1,w_2, ...,w_{c-1},w_c]\)

\subsubsection{Multi-Class Hinge Loss}
Hinge loss function was initially proposed as an alternative to cross-entropy for binary classification. It was primarily developed for use with SVM models. The targets values are in the set \{–1, 1\}. The main purpose of hinge loss is to maximise the decision boundary between two groups that are to be discriminated, that is, a binary classification problem. For linear classifiers {\textit{w,x}}, hinge loss is computed as follows:
\begin{equation}{L(w,(x,y))} = max\{0,1-y\{w,x\}\} \end{equation}

Squared hinge loss function was developed as an extension, which computes the square of the score hinge loss. It smoothens the surface of the error function, thus making it easier for computational purposes. 

Categorical hinge loss or multi-class hinge loss is computed as follows, "For a prediction y, take all values unequal to t, and compute the loss. Eventually, sum them all together to find the multi-class hinge loss"
\cite{Weston1999SupportVM,ZHANG201455,Alexander-lecture05,shavlev-et-al}. The regular targets are computed into categorical data, and the loss function is calculated as follows:
\begin{equation}l(y) = \sum_{y\neq t}max(0,1 + W_yX - W_t X)\end{equation}

%Your text comes here. Separate text sections with
%\section{Section title}
%\label{sec:1}
%Text with citations \cite{RefB} and \cite{RefJ}.
%\subsection{Subsection title}
%\label{sec:2}
%as required. Don't forget to give each section
%and subsection a unique label (see Sect.~\ref{sec:1}).
%\paragraph{Paragraph headings} Use paragraph headings as needed.
%\begin{equation}
%a^2+b^2=c^2
%\end{equation}

\begin{table*}[ht]

\begin{tabular}{l|rrrr|rrrr}
\hline
Models & BERT & DistilBERT & ALBERT  & RoBERTa  & XLM & XLNet & XLM-R & CharacterBERT\\ \hline
 
\multicolumn{9}{c}{Precision (sentiment analysis)}    \\ \hline
{Positive}                      & {0.701}     & {0.692}           & {0.634}       & {0.667}        & {0.681}    & {0.711}      & {0.567}      & {0.631}         \\
{Negative}                      & {0.573}     & {0.501}           & {0.556}       & {0.120}        & {0.552}    & {0.571}      & {0.520}      & {0.689}         \\
{Mixed feelings}                & {0.329}     & {0.443}           & {0.0}       & {0.162}        & {0.266}    & {0.292}      & {0.0}      & {0.248}         \\
{Neutral}                       & {0.371}     & {0.551}           & {0.350}       & {0.264 }        & {0.568}    & {0.498}      & {0.467}      & {0.602}         \\
{Other language}                & {0.564}     & {0.192}           & {0.496}       & { 0.012 }        & {0.498}    & {0.508}      & {0.073}      & {0.503}         \\ \hline
{Macro-average}      & \textbf{0.509}     & {0.471}           & {0.427}       & {0.245}        & {0.461}    & {0.441}      & {0.325}      & {0.452}         \\
{Weighted average}                & \textbf{0.592}     & {0.564}           & {0.536}       & {0.362}        & {0.519}    & {0.587}      & {0.418}      & {0.542}         \\ \hline
\multicolumn{9}{ c}{Precision (Offensive language identification)}                                                                                                                                                                                                                                   \\ \hline
{Not offensive}                 & {0.781}     & {0.778}           & {0.715}       & {0.679}        & {0.759}    & {0.719}      & {0.735}      & {0.749}         \\
{Offensive untargeted}          & {0.0}     & {0.123}           & {0.0}       & {0.0}        & {0.0}    & {0.0}      & {0.0}      & {0.00}         \\
{Offensive targeted individual} & {0.452}     & {0.461}           & {0.654}       & {0.769}        & {0.798}    & {0.641}      & {0.085}      & {0.721}         \\
{Offensive targeted group}      & {0.0}     & {0.391}           & {0.0}       & {0.571}        & {0.371}    & {0.0}      & {0.0}      & {0.428}         \\
{Offensive targeted others}      & {0.0}     & {0.0}           & {0.0}       & {0.0}        & {0.0}    & {0.0}      & {0.0}      & {0.0}         \\
{Other languages}                & {0.704}     & {0.731}           & {0.702}       & {0.748}        & {0.678}    & {0.712}      & {0.008}      & {0.678}         \\ \hline   
{Macro-average}      & {0.321}     & {0.408}           & {0.345}       & \textbf {0.461}        & {0.431}    & {0.345}      & {0.138}      & {0.431}         \\
{Weighted average}                & {0.651}     & \textbf {0.691}           & {0.622}       & {0.656}        & {0.678}    & {0.625}      & {0.425}      & {0.669}         \\ \hline
 
\multicolumn{9}{ c}{{Recall (sentiment analysis)}}                                                                                                                                                                                                                                              \\ \hline
{Positive}                      & {0.742}     & {0.753}           & {0.793}       & {0.667}        & {0.663} & {0.649}      & {0.759}      & {0.867}         \\
{Negative}                      & {0.581}     & {0.442}           & {0.524}       & {0.467}        & {0.676}    & {0.778}      & {0.604}      & {0.532}         \\
{Mixed feelings}                & {0.100}     & {0.229}           & {0.0}       & {0.074}        & {0.108}    & {0.061}      & {0.0}      & {0.051}         \\
{Neutral}                       & {0.493}     & {0.581}           & {0.263}       & {0.144}        & {0.376}    & {0.401}      & {0.073}      & {0.343}         \\
{Other language}                & {0.590}     & {0.209}           & {0.678}       & {0.006}        & {0.608}    & {0.702}      & {0.117}      & {0.618}         \\ \hline
{Macro-average}      & {0.501}     & {0.441}           & {0.452}       & {0.271}        & {0.508}    & \textbf {0.521}      & {0.311}      & {0.479}         \\
{Weighted average}                & {0.602}     & {0.572}           & {0.592}       & {0.361}        & \textbf {0.628}    & {0.612}      & {0.470 }      & {0.628}         \\ \hline
\multicolumn{9}{ c}{{Recall (Offensive language identification)}}                                                                                                                                                                                                                                   \\ \hline
{Not offensive}                 & {0.842}     & {0.813}           & {0.875}       & {0.917}        & {0.858}    & {0.708}      & {0.746}      & {0.864}         \\
{Offensive untargeted}          & {0.0}     & {0.041}           & {0.0}       & {0.0}        & {0.0}    & {0.0}      & {0.0}      & {0.0}         \\
{Offensive targeted individual} & {0.533}     & {0.629}           & {0.415}       & {0.484 }        & {0.599}    & {0.610}      & {0.025}      & {0.653}         \\
{Offensive targeted group}      & {0.0}     & {0.251}           & {0.0}       & {0.089}        & {0.229}    & {0.0}      & {0.0}      & {0.303}         \\
{Offensive targeted others}      & {0.0}     & {0.0}           & {0.0}       & {0.0}        & {0.0}    & {0.0}      & {0.0}      & {0.0}         \\
{Other languages}                & {0.741}     & {0.748}           & {0.812}       & { 0.601 }        & {0.768}    & {0.872}      & {0.038}      & {0.731}         \\ \hline  
{Macro-average}      & {0.379}     & \textbf {0.412}           & {0.350}       & {0.349}        & {0.408}    & {0.365}      & {0.135}      & {0.409}         \\
{Weighted average}                & \textbf{0.719}     & {0.709}           & {0.707}       & {0.696}        & {0.714}    & {0.706}      & {0.418}      & {0.718}         \\ \hline  
 
\multicolumn{9}{ c}{{F1-score (sentiment analysis)}}                                                                                                                                                                                                                                              \\ \hline
{Positive}                      & {0.721}     & {0.723}           & {0.705}       & {0.665 }        & {0.714}    & {0.681}      & {0.649}      & {0.729}         \\
{Negative}                      & {0.573}     & {0.473}           & {0.583}       & {0.190}        & {0.662}    & {0.663}      & {0.559}      & {0.608}         \\
{Mixed feelings}                & {0.153}     & {0.301}           & {0.0}       & {0.102}        & {0.182}    & {0.102}      & {0.0}      & {0.089}         \\
{Neutral}                       & {0.424}     & {0.558}           & {0.300}       & {0.187}        & {.449}    & {0.443}      & {0.126}      & {0.431}         \\
{Other languages}                & {0.581}     & {0.201}           & {0.573}       & {0.008}        & {0.548}    & {0.591}      & {0.090}      & {0.609}         \\ \hline
{Macro-average}      & {0.493}     & {0.453}           & {0.432}       & {0.230}        & \textbf{0.498}    & {0.491}      & {0.285}      & {0.489}         \\
{Weighted average}                & \textbf{0.591}     & {0.561}           & {0.556}       & {0.349}        & {0.595}    & {0.589}      & {0.418}      & {0.594}         \\ \hline
\multicolumn{9}{ c}{{F1-score (offensive language identification)}}                                                                                                                                                                                                                                   \\ \hline
{Not offensive}                 & {0.811}     & {0.801}           & {0.787}       & {0.780}        & {0.792}    & {0.788}      & {0.741}      & {0.801}         \\
{Offensive untargeted}          & {0.0}     & {0.061}           & {0.0}       & {0.0}        & {0.0}    & {0.0}      & {0.0}      & {0.0}         \\
{Offensive targeted individual} & {0.529}     & {0.534}           & {0.507}       & {0.594}        & {0.685}    & {0.625}      & {0.039}      & {0.649}         \\
{Offensive targeted group}      & {0.0}     & {0.299}           & {0.0}       & {0.154}        & {0.285}    & {0.0}      & {0.0}      & {0.348}         \\
{Offensive targeted others}      & {0.0}     & {0.0}           & {0.0}       & {0.0}        & {0.0}    & {0.0}      & {0.
0}      & {0.0}         \\
{Other languages}                & {0.743}     & {0.741}           & {0.753}       & {0.667}        & {0.725}    & {0.710}      & {0.013}      & {0.698}         \\ \hline   
{Macro-average}      & \textbf{0.351}     & {0.421}           & {0.341}       & {0.366}        & {0.419}    & {0.354}      & {0.132}      & {0.418}         \\
{Weighted average}                & {0.686}     & \textbf{0.698}           & {0.656}       & {0.651}        & {0.685}    & {0.661}      & {0.416}      & {0.680}         \\ \hline
                                                    
\end{tabular}
 \caption{Precision, recall, and F1-scores of STL approach for Kannada} \label{stl_kan}
\end{table*}

\subsubsection{Focal Loss}
The focal loss was initially proposed for dense object detection task as an alternative to CE loss. It enables training highly accurate dense object detectors for highly imbalanced datasets \cite{Lin_2017_ICCV}. For imbalanced text datasets, it handles class imbalance for by calculating the ratio of each class and thus assigning weights based on whether it is hard or soft examples (inliers) \cite{tula-etal-2021-bitions,ma-etal-2020-xlp}.
\begin{table*}[ht]
 
\begin{tabular}{l|rrrr|rrrr}
\hline
Models & BERT & DistilBERT & ALBERT  & RoBERTa  & XLM & XLNet & XLM-R & CharacterBERT\\ \hline
 
\multicolumn{9}{c}{Precision (sentiment analysis)}                                                                                                                                                                                                                                                  \\ \hline
{Positive}                      & {0.704}     & {0.657}           & {0.446}       & {0.615}        & {0.713}    & {0.695}      & { 0.750}      & {0.673}         \\
{Negative}                      & {0.343}     & {0.0}           & {0.0}       & {0.512}        & {0.536}    & {0.521}      & {0.494}      & {0.623}         \\
{Mixed feelings}                & {0.0}     & {0.0}           & {0.0}       & {0.175}        & {0.521}    & {0.571}      & { 0.562}      & {0.447}         \\
{Neutral}                       & {0.664}     & {0.600}           & {0.333}       & {0.712 }        & {0.765}    & {0.591}      & {0.751}      & {0.679}         \\
{Other languages}                & {0.750}     & {0.750}           & {0.622}       & {0.744}        & {0.735}    & {0.814}      & {0.706 }      & {0.787}         \\ \hline

{Macro-average}      & {0.492}     & {0.402}           & {0.280}       & { 0.552}        & {0.654}    & {0.639}      & {\textbf{0.653}}      & {0.642}         \\ 
{weighted aAverage}                & {0.612}     & {0.533}           & {0.346}       & {0.618}        & {0.701}    & {0.643}      & {\textbf{0.708}}      & {0.664}         \\ \hline
 
\multicolumn{9}{ c}{Precision (Offensive language identification)}                                                                                                                                                                                                                                   \\ \hline
{Not offensive}                 & {0.939}     & {0.947}           & {0.933}       & {0.946}        & {0.929}    & {0.951}      & {0.716}      & {0.934}         \\
{Offensive untargeted}          & {0.0}     & {0.529}           & {0.0}       & {0.400}        & {0.0}    & {0.0}      & {0.0}      & {1.000}         \\
{Offensive targeted individual} & {0.0}     & {0.222}           & {0.0}       & {0.0}        & {0.0}    & {0.027}      & {0.0}      & {0.0}         \\
{Offensive targeted group}      & {0.0}     & {0.0}           & {0.0}       & {0.0}        & {0.0}    & {0.028}      & {0.0}      & {0.0}         \\ 
{Other languages}                & {0.717}     & {0.716}           & {0.744}       & {0.807}        & {0.756}    & {0.045}      & {0.829}      & {0.724}         \\ \hline

{Macro-average}      & {0.331}     & {\textbf{0.483}}           & {0.335}       & {0.431}        & {0.337}    & {0.210}      & {0.353}      & {0.532}         \\ 
{weighted average}                & {0.884}     & {\textbf{0.905}}           & {0.880}       & {0.904}        & {0.878}    & {0.847}      & {0.885}      & {0.901}         \\ \hline
 
\multicolumn{9}{ c}{ Recall (sentiment analysis)}                                                                                                                                                                                                                                              \\ \hline
{Positive}                      & {0.801}     & {0.826}           & {0.963}       & {0.829}        & {0.827}    & {0.697}      & {0.776}      & {0.814}         \\
{Negative}                      & {0.274}     & {0.0}           & {0.0}       & {0.304}        & {0.437}    & {0.363}      & {0.630}      & {0.244}         \\
{Mixed feelings}                & {0.0}     & {0.0}           & {0.0}       & {0.136}        & {0.309}    & {0.147}      & {0.233}      & {0.159}         \\
{Neutral}                       & {0.719}     & {0.717}           & {0.009}       & {0.531}        & {0.505}    & {0.559}      & {0.596}      & {0.712}         \\
{Other languages}                & {0.742}     & {0.717}           & {0.596}       & {0.681}        & {0.566}    & {0.411}      & {0.566}      & {0.528}         \\ \hline

{Macro-average}      & \textbf{0.506}     & {0.452}           & {0.314}       & {0.496}        & {0.509}    & {0.495}      & {{0.503}}      & {0.502}         \\ 
{weighted average}                & \textbf{0.663}     & {0.642}           & {0.457}       & {0.620}        & {0.608}    & {0.640}      & {{0.609}}      & {0.621}         \\ \hline
 
\multicolumn{9}{ c}{Recall (Offensive language identification)}                                                                                                                                                                                                                                   \\ \hline
{Not offensive}                 & {0.983}     & {0.966 }           & {0.981}       & {0.976}        & {0.983}    & {0.648}      & {0.935}      & {0.979}         \\
{Offensive untargeted}          & {0.0}     & {0.321}           & {0.0}       & {0.286 }        & {0.0}    & {0.0}      & {0.0}      & {0.071}         \\
{Offensive targeted individual} & {0.0}     & {0.125}           & {0.0}       & {0.0}        & {0.0}    & {0.125}      & {0.0}      & {0.0}         \\
{Offensive targeted group}      & {0.0}     & {0.0}           & {0.0}       & {0.0}        & {0.0}    & {0.100}      & {0.0}      & {0.0}         \\ 
{Other languages}                & {0.789}     & {0.756}           & {0.711}       & {0.789}        & {0.656}    & {0.156}      & {0.815}      & {0.700}         \\ \hline

{Macro-average}      & {0.353}     & {\textbf{0.434}}           & {0.338}       & {0.410}        & {0.328}    & {0.206}      & {0.359}      & {0.350}         \\ 
{weighted average}                & {0.922}     & {0.919}           & {0.920}       & {\textbf{0.928}}        & {0.919}    & {0.588}      & {0.886}      & {0.919}         \\ \hline
 
\multicolumn{9}{ c}{F1-score (sentiment analysis)}                                                                                                                                                                                                                                              \\ \hline
{Positive}                      & {0.750}     & {0.732}           & {0.609}       & {0.706}        & {0.666}    & {0.666}      & {0.662}      & {0.637}         \\
{Negative}                      & {0.305}     & {0.0}           & {0.0}       & {0.381}        & {0.482}    & {0.428}      & {0.554}      & {0.351}         \\
{Mixed feelings}                & {0.0}     & {0.0}           & {0.0}       & {0.153}        & {0.388}    & {0.145}      & {0.219}      & {0.128}         \\
{Neutral}                       & {0.691}     & {0.654}           & {0.018}       & {0.608}        & {0.634}    & {0.665}      & {0.622}      & {0.655}         \\
{Other languages}                & {0.742}     & {0.733}           & {0.609}       & {0.711}        & {0.591}    & {0.627}      & {0.605}      & {0.598}         \\ \hline
{Macro-average}      & \textbf{0.497}     & {0.424}           & {0.247}       & {0.512}        & {0.624}    & {0.481}      & \textbf{}{{0.488}}      & {0.492}         \\ 
{weighted average}                & \textbf{0.635}     & {0.581}           & {0.310}       & {0.605}        & {0.623}    & {0.630}      & {{0.603}}      & {0.624}         \\ \hline
 
\multicolumn{9}{ c}{{F1-score (offensive language identification)}}                                                                                                                                                                                                                                   \\ \hline
{Not offensive}                 & {0.958}     & {0.956}           & {0.956}       & {0.961}        & {0.955}    & {0.771}      & {0.962}      & {0.956}         \\
{Offensive untargeted}          & {0.0}     & {0.400}           & {0.0}       & {0.333}        & {0.0}    & {0.0}      & {0.0}      & {0.133}         \\
{Offensive targeted individual} & {0.0}     & {0.160}           & {0.0}       & {0.0}        & {0.0}    & {0.044}      & {0.0}      & {0.0}         \\
{Offensive targeted group}      & {0.0}     & {0.0}           & {0.0}       & {0.0}        & {0.0}    & {0.043}      & {0.0}      & {0.0}         \\ 
{Other languages}                & {0.751}     & {0.685 }           & {0.727}       & {0.798}        & {0.702}    & {0.070}      & {0.836}      & {0.712}         \\ \hline

{Macro-average}      & {0.342}     & {\textbf{0.440}}           & {0.337}       & {0.418}        & {0.332}    & {0.186}      & {0.356}      & {0.360}         \\ 
{weighted average}                & \textbf{0.902}     & {0.901}           & {0.900}       & {0.900}        & {0.897}    & {0.690}      & { {0.898}}      & {0.901}         \\ \hline

\end{tabular}
\caption{Precision, recall, and F1-scores of STL approach for Malayalam} \label{stl_mal}
\end{table*}
In CE, easily classified examples incur a loss with non-trivial magnitude. We define focal loss by adding a modulating factor loss to the cross-entropy loss with a tunable focusing parameter \(\gamma\geq0\) \cite{Lin_2017_ICCV} as:
\begin{equation}FL({p_t}) = -(1 - {p_t})^{\gamma}\log({p_t})\end{equation}
where \(\emph{p} \in [0,1]\) is the estimated probability of the model for a given class. When \(\gamma = 0\), the above formula represents CE. If \(\gamma > 1\), the modulating factor assists in reducing the loss for easily classified examples (\(p_t\geq0.5\)) resulting in more number of corrections of misclassified examples.
\subsubsection{Kullback–Leibler Divergence Loss}

The Kullback–Leibler divergence (KLD) measures the difference in the probability distribution of two random variables \cite{kullback1951}. Theoretically, a KLD score of 0 indicates that the two distributions are identical.
\begin{table*}[htbp]
 
\begin{tabular}{l|rrrr|rrrr}
\hline 
Models & BERT & DistilBERT & ALBERT  & RoBERTa  & XLM & XLNet & XLM-R & CharacterBERT\\ \hline
 
\multicolumn{9}{c}{Precision (sentiment analysis)}                                                                                                                                                                                                                                              \\ \hline
{Positive}                      & {0.713}     & {0.740}           & {0.664}       & {0.702}        & {0.689}    & {0.686}      & {0.714}      & {0.676}         \\
{Negative}                      & {0.413}     & {0.417}           & {0.393}       & {0.392}        & {0.462}    & {0.444}      & {0.399}      & {0.505}         \\
{Mixed feelings}                & {0.411}     & {0.329}           & {0.436}       & {0.345}        & {0.440}    & {0.419}      & {0.397}      & {0.307}         \\
{Neutral}                       & {0.444}     & {0.472}           & {0.464}       & {0.465}        & {0.534}    & {0.423}      & {0.526}      & {0.540}         \\
{Other language}                & {0.713}     & {0.621}           & {0.667}       & {0.630}        & {0.675}    & {0.660}      & {0.517}      & {0.543}         \\ \hline

{Macro-average}                       & {0.520}     & {0.516}           & {0.525}       & {0.507}        & {\textbf{0.560}}    & {0.527}      & {0.511}      & {0.514}         \\
{Weighted average}                & {0.596}     & {0.607}           & {0.574}       & {0.584}        & {\textbf{ 0.609}}    & {0.584}      & {0.602}      & {0.587}         \\ \hline
\multicolumn{9}{ c}{Precision (Offensive language identification)}                                                                                                                                                                                                                                   \\ \hline
{Not offensive}                 & {0.797}     & {0.860}           & {0.796}       & {0.826}        & {0.815}    & {0.837}      & {0.853}      & {0.839}         \\
{Offensive untargeted}          & {0.294}     & {0.385}           & {0.284}       & {0.438}        & {0.428}    & {0.370}      & {0.409}      & {0.368}         \\
{Offensive targeted individual} & {0.0}     & {0.371}           & {0.0}       & {0.384}        & {0.382}    & {0.377}      & {0.409}      & {0.349}         \\
{Offensive targeted group}      & {0.0}     & {0.318}           & {0.0}       & {0.297}        & {0.423}    & {0.329}      & {0.454}      & {0.323}         \\
{Offensive targeted others}      & {0.0}     & {0.0}           & {0.0}       & {0.0}        & {0.0}    & {0.0}      & {0.0}      & {0.0}         \\
{Other language}                & {0.840}     & {0.829}           & {0.727}       & {0.857}        & {0.780}    & {0.660}      & {0.806}      & {0.780}         \\ \hline

{Macro-average}                       & {0.322}     & \textbf{0.461}           & {0.301}       & {0.467}        & {0.472}    & {0.429}      & {{0.458}}      & {0.451}         \\
{Weighted average}                & {0.636}     & \textbf{0.738}           & {0.630}       & {0.718}        & {0.715}    & {0.714}      & {{0.725}}      & {0.719}         \\ \hline

\multicolumn{9}{ c}{Recall (sentiment analysis)}                                                                                                                                                                                                                                              \\ \hline
{Positive}                      & {0.854}     & {0.805}           & {0.882}       & {0.827}        & {0.888}    & {0.863}      & {0.831}      & {0.891}         \\
{Negative}                      & {0.391}     & {0.401}           & {0.265}       & {0.427}        & {0.329}    & {0.323}      & {0.440}      & {0.274}         \\
{Mixed feelings}                & {0.129}     & {0.228}           & {0.131}       & {0.123}        & {0.174}    & {0.095}      & {0.174}      & {0.144}         \\
{Neutral}                       & {0.373}     & {0.442}           & {0.322}       & {0.365}        & {0.366}    & {0.386}      & {0.363}      & {0.298}         \\
{Other language}                & {0.554}     & {0.601}           & {0.404}       & {0.545}        & {0.526}    & {0.474}      & {0.582}      & {0.620}         \\ \hline

{Macro-average}                       & {0.460}     & \textbf{0.495}           & {0.401}       & {0.457}        & {0.457}    & {0.428}      & {{0.478}}      & {0.445}         \\
{Weighted average}                & {0.627}     & {0.625}           & {0.612}       & {0.614}        & {\textbf{0.641}}    & {0.618}      & {0.625}      & {0.626}         \\ \hline
\multicolumn{9}{ c}{Recall (Offensive language identification)}                                                                                                                                                                                                                                   \\ \hline
{Not offensive}                 & {0.964}     & {0.887}           & {0.947}       & {0.935}        & {0.946}    & {0.910}      & {0.915}      & {0.919}         \\
{Offensive untargeted}          & {0.345}     & {0.408}           & {0.384}       & {0.339}        & {0.432}    & {0.426}      & {0.479}      & {0.414}         \\
{Offensive targeted individual} & {0.0}     & {0.288}           & {0.0}       & {0.266}        & {0.082}    & {0.275}      & {0.326}      & {0.288}         \\
{Offensive targeted group}      & {0.0}     & {0.351}           & {0.0}       & {0.134}        & {0.154}    & {0.092}      & {0.210}      & {0.102}         \\
{Offensive targeted others}      & {0.0}     & {0.0}           & {0.0}       & {0.0}        & {0.0}    & {0.0}      & {0.0}      & {0.0}         \\
{Other languages}                & {0.618}     & {0.708}           & {0.539}       & {0.742}        & {0.719}    & {0.775}      & {0.715}      & {0.685}         \\ \hline

{Macro-average}                       & {0.321}     & \textbf{0.440}           & {0.312}       & {0.403}        & {0.389}    & {0.413}      & { {0.439}}      & {0.401}         \\
{Weighted average}                & {0.752}     & {0.750}           & {0.740}       & {0.765}        & \textbf{0.767}    & {0.752}      & {{0.741}}      & {0.755}         \\ \hline
 
\multicolumn{9}{ c}{F1-score (sentiment analysis)}                                                                                                                                                                                                                                              \\ \hline
{Positive}                      & {0.777}     & {0.771}           & {0.758}       & {0.759}        & {0.776}    & {0.765}      & {0.768}      & {0.769}         \\
{Negative}                      & {0.402}     & {0.409}           & {0.316}       & {0.409}        & {0.384}    & {0.374}      & {0.419}      & {0.355}         \\
{Mixed feelings}                & {0.196}     & {0.269}           & {0.202}       & {0.181}        & {0.250}    & {0.154}      & {0.242}      & {0.196}         \\
{Neutral}                       & {0.406}     & {0.456}           & {0.380}       & {0.409}        & {0.434}    & {0.404}      & {0.430}      & {0.384}         \\
{Other languages}                & {0.586}     & {0.611}           & {0.503}       & {0.584}        & {0.591}    & {0.552}      & {0.552}      & {0.579}         \\ \hline

{Macro-average}                       & {0.473}     & \textbf{0.503}           & {0.432}       & {0.469}        & {0.487}    & {0.450}      & {{0.489}}      & {0.457}         \\
{Weighted average}                & {0.600}     & \textbf{0.614}           & {0.571}       & {0.589}        & {0.607}    & {0.583}      & {{0.609}}      & {0.585}         \\ \hline
\multicolumn{9}{ c}{F1-score (offensive language identification)}                                                                                                                                                                                                                                   \\ \hline
{Not offensive}                 & {0.873}     & {0.873}           & {0.865}       & {0.877}        & {0.876}    & {0.872}      & {0.882}      & {0.877}         \\
{Offensive untargeted}          & {0.317}     & {0.396}           & {0.326}       & {0.383}        & {0.430}    & {0.396}      & {0.441}      & {0.387}         \\
{Offensive targeted individual} & {0.0}     & {0.324}           & {0.0}       & {0.314}        & {0.135}    & {0.318}      & {0.362}      & {0.315}         \\
{Offensive targeted group}      & {0.0}     & {0.334}           & {0.0}       & {0.185}        & {0.226}    & {0.144}      & {0.287}      & {0.155}         \\
{Offensive targeted others}      & {0.0}     & {0.0}           & {0.0}       & {0.0}        & {0.0}    & {0.0}      & {0.0}      & {0.0}         \\
{Other languages}                & {0.712}     & {0.764}           & {0.619}       & {0.795}        & {0.749}    & {0.713}      & {0.737}      & {0.751}         \\ \hline

{Macro-average}                       & {0.317}     & \textbf{0.448}           & {0.302}       & {0.426}        & {0.403}    & {0.407}      & {{0.430}}      & {0.414}         \\
{Weighted average}                & {0.688}     & \textbf{0.743}           & {0.679}       & {0.735}        & {0.726}    & {0.727}      & {{0.734}}      & {0.731}         \\ \hline

\end{tabular}
\caption{Precision, recall, and F1-scores of STL approach for Tamil}
\label{stl_tam}
\end{table*}

\begin{table*}[htbp]
 
\begin{tabular}{l|rrrr|rrrr}
\hline
\multicolumn{1}{c|}{} & \multicolumn{4}{c|}{Hard parameter sharing} & \multicolumn{4}{c}{Soft parameter sharing} \\ \hline
Losses & Cross-entropy  &   Hinge loss &  Focal loss    &  KLD    & CE &    HL  &   FL   &  KLD\\ \hline
\multicolumn{9}{c}{Precision (sentiment analysis)}\\  \hline                                    
{Positive}                      & {0.683}     & {0.673}           & {0.666}       & {0.459}        & {0.713}    & {0.534}      & {0.679}      & {0.459}         \\
{Negative}                      & {0.677}    & {0.685}           & {0.651}       & {0.0}        & {0.596}    & {0.733}      & {0.631}      & {0.0}         \\
{Mixed feelings}                & {0.0}     & {0.0}           & {0.0}       & {0.0}        & {0.143}    & {0.0}      & {0.0}      & {0.0}         \\
{Neutral}                 & {0.5}     & {0.700}           & {0.479}       & {0.0}        & {0.491}    & {0.577}      & {0.492}      & {0.0}         \\
{Other languages}                & {0.442}     & {0.477}           & {0.459}       & {0.0}        & {0.538}    & {0.703}      & {0.454}      & {0.0}         \\ \hline

{Macro-average}                       & {0.460}     & \textbf{0.507}           & {0.451}       & {0.092}        & { {0.496}}    & {0.509}      & {0.451}      & {0.092}         \\
{Weighted average}                & {0.574}     & \textbf{0.597}           & {0.560}       & {0.210}        & { { 0.592}}    & {0.558}      & {0.562}      & {0.210}      \\ \hline
\multicolumn{9}{ c}{Precision (Offensive language identification)}                                                                                                                                                                                                                                   \\ \hline
{Not offensive}                 & {0.802}     & {0.755}           & {0.764}       & {0.559}        & {0.727}    & {0.707}      & {0.796}      & {0.559}         \\
{Offensive untargeted}          & {0.0}     & {0.0}           & {0.0}       & {0.0}        & {0.0}    & {0.0}      & {0.0}      & {0.0}         \\
{Offensive targeted individual} & {0.640}     & {0.625}           & {0.588}       & {0.0}        & {0.641}    & {0.838}      & {0.718}      & {0.0}         \\
{Offensive targeted group}      & {0.171}     & {0.0}           & {0.0}       & {0.0}        & {0.250}    & {0.0}      & {0.132}      & {0.0}         \\
{Offensive targeted others}      & {0.0}     & {0.0}           & {0.0}       & {0.0}        & {0.0}    & {0.0}      & {0.0}      & {0.0}         \\
{Other languages}                & {0.703}     & {0.700}           & {0.709}       & {0.0}        & {0.730}    & {0.716}      & {0.682}      & {0.0}         \\ \hline

{Macro-average}                       & \textbf{0.680}     & {0.348}           & {0.344}       & {0.093}        & {0.391}    & {0.377}      & { {0.388}}      & {0.093}         \\
{Weighted average}                & \textbf{0.680}     & {0.642}           & {0.644}       & {0.313}        & {0.648}    & {0.641}      & { {0.678}}      & {0.313}         \\ \hline

\multicolumn{9}{ c}{ Recall (sentiment analysis)}                                                                                                                                                                                                                                              \\ \hline
{Positive}                      & {0.787}     & {0.790}           & {0.781}       & {1.0}        & {0.746}    & {0.967}      & {0.772}      & {1.0}         \\
{Negative}                      & {0.689}     & {0.689}           & {0.659}       & {0.0}        & {0.774}    & {0.268}      & {0.689}      & {0.0}         \\
{Mixed feelings}                & {0.0}     & {0.0}           & {0.0}       & {0.0}        & {0.016}    & {0.0}      & {0.0}      & {0.0}         \\
{Unknown state}                       & {0.237}     & {0.287}           & {0.322}       & {0.0}        & {0.325}    & {0.188}      & {0.375}      & {0.0}         \\
{Other languages}                & {0.701}     & {0.828}           & {0.644}       & {0.0}        & {0.655}    & {0.299}      & {0.563}      & {0.0}         \\ \hline

{Macro-average}                       & {0.483}     & {0.496}           & {0.474}       & {0.200}        & \textbf{0.503}    & {0.344}      & { {0.480}}      & {0.200}         \\
{Weighted average}                & \textbf{0.626}     &  {0.636}           & {0.615}       & {0.459}        & { {0.632}}    & {0.560}      & {0.618}      & {0.459}         \\ \hline
\multicolumn{9}{c}{ Recall (Offensive language identification)}                                                                                                                                                                                                                                   \\ \hline
{Not offensive}                 & {0.828}     & {0.857}           & {0.853}       & {1.0}        & {0.882}    & {0.870}      & {0.794}      & {1.0}         \\
{Offensive untargeted}          & {0.0}     & {0.0}           & {0.0}       & {0.0}        & {0.0}    & {0.0}      & {0.0}      & {0.0}         \\
{Offensive targeted individual} & {0.585}     & {0.671}           & {0.610}       & {0.0}        & {0.610}    & {0.378}      & {0.622}      & {0.0}         \\
{Offensive targeted group}      & {0.159}     & {0.0}           & {0.0}       & {0.0}        & {0.023}    & {0.0}      & {0.159}      & {0.0}         \\
{Offensive targeted others}      & {0.0}    & {0.0}           & {0.0}       & {0.0}        & {0.0}    & {0.0}      & {0.0}      & {0.0}         \\
{Other languages}                & {0.877}     & {0.818}           & {0.870}       & {0.0}        & {0.721}    & {0.883}      & {0.877}      & {0.0}         \\ \hline

{Macro-average}                       & {0.408}     & {0.391}           & {0.389}       & {0.167}        & {0.373}    & {0.355}      & \textbf{ {0.409}}      & {0.167}         \\
{Weighted average}                & {0.724}     & \textbf{0.728}           & {0.729}       & {0.559}        & {0.716}    & {0.716}      & { {0.709}}      & {0.559}         \\ \hline

\multicolumn{9}{ c}{F1-score (sentiment analysis)}                                                                                                                                                                                                                                              \\ \hline
{Positive}                      & {0.732}     & {0.727}           & {0.719}       & {0.629}        & {0.729}    & {0.688}      & {0.723}      & {0.629}         \\
{Negative}                      & {0.683}     & {0.687}           & {0.655}       & {0.0}        & {0.674}    & {0.393}      & {0.659}      & {0.0}         \\
{Mixed feelings}                & {0.0}     & {0.0}           & {0.0}       & {0.0}        & {0.029}    & {0.0}     & {0.0}      & {0.0}         \\
{Neutral}                       & {0.322}     & {0.280}           & {0.359}       & {0.0}        & {0.391}    & {0.283}      & {0.426}      & {0.0}         \\
{Other languages}                & {0.542}     & {0.605}           & {0.536}       & {0.0}        & {0.591}    & {0.419}      & {0.603}      & {0.0}         \\ \hline

{Macro-average}                       & {0.456}     & {0.460}           & {0.454}       & {0.126}        & {0.483}    & {0.357}      & \textbf{ {0.462}}      & {0.126}         \\
{Weighted average}                & {0.590}     & {0.591}           & {0.581}       & {0.289}        & {0.602}    & {0.485}      & { {0.587}}      & 0.289         \\ \hline
\multicolumn{9}{ c}{F1-score (offensive language identification)}                                       \\ \hline
{Not offensive}                 & {0.815}     & {0.803}           & {0.806}       & {0.717}        & {0.797}    & {0.780}      & {0.795}      & {0.717}         \\
{Offensive untargeted}          & {0.0}     & {0.0}           & {0.0}       & {0.0}        & {0.0}    & {0.0}      & {0.0}      & {0.0}         \\
{Offensive targeted individual} & {0.611}     & {0.647}           & {0.610}       & {0.0}        & {0.625}    & {0.521}      & {0.667}      & {0.0}         \\
{Offensive targeted group}      & {0.165}     & {0.0}           & {0.0}       & {0.0}        & {0.042}    & {0.0}      & {0.144}      & {0.0}         \\
{Offensive targeted others}      & {0.0}     & {0.0}           & {0.0}       & {0.0}        & {0.0}    & {0.0}      & {0.0}      & {0.0}         \\
{Other languages}                & {0.780}     & {0.759}           & {0.781}       & {0.0}        & {0.725}    & {0.791}      & {0.767}      & {0.0}         \\ \hline

{Macro-average}                       & \textbf{0.395}     & {0.368}           & {0.364}       & {0.120}        & {0.365}    & {0.349}      & \textbf{{0.395}}      & {0.120}         \\
{Weighted average}                & \textbf{0.700}     & {0.683}           & {0.683}       & {0.401}        & {0.672}    & {0.662}      & {{0.690}}      & {0.401}         \\ \hline

\end{tabular}
\caption{MTL results on the Kannada data set}\label{mtl_kan}
\end{table*}

It is defined as "the average number of extra bits needed to encode the data, due to the fact that we used the distribution q to encode the data instead of the true distribution p" \cite{MurphyML} and also known as relative entropy. It is generally used for models that have complex functions such as variational autoencoders (VAEs) for text generation rather than simple multi-class classification tasks \cite{prokhorov-etal-2019-importance,9244048}. Consider two random distributions, P and Q. The main intuition behind KLD is that if the probability for an event from Q is small, but that of P is large; then there exists a large divergence. It is used to compute the divergence between discrete and continuous probability distributions \cite{KLD_ml_mastery}. It is computed as follows:
\begin{equation}D_{KL}(p||q) = \sum_{i=1}^N p(x_i)\log\frac{p(x_i)}{q(x_i)}\end{equation}

% For one-column wide figures use
 
%
% For two-column wide figures use
 
% For tables use
%\begin{table}
% table caption is above the table
%\caption{Please write your table caption here}
%\label{tab:1}       % Give a unique label
% For LaTeX tables use
%\begin{tabular}{lll}
%\hline\noalign{\smallskip}
%first & second & third  \\
%\noalign{\smallskip}\hline\noalign{\smallskip}
%number & number & number \\
%number & number & number \\
%\noalign{\smallskip}\hline
%\end{tabular}
%\end{table}
\section{Experiments}
\subsection{Experiments Setup}
We use the pretrained models available on huggingface transformers\footnote{{https://huggingface.co/transformers/index.html}} \cite{wolf-etal-2020-transformers}, while fine-tuning the model on a python-based deep learning framework, Pytorch\footnote{\url{https://pytorch.org/}} \cite{Paszke2019PyTorchAI} and Scikit-Learn\footnote{\url{https://scikit-learn.org/}} \cite{Pedregosa2011ScikitlearnML} to evaluate the performance of the models. For MTL experiments, we take the naive weighted sum by assigning equal weights to the losses \cite{7410661}. 
We have experimented with various hyperparameters that are listed as shown in Table\,\ref{params}. The experiments were conducted by dividing the data set into three parts: 80\% for training, 10\% for validation, and 10\% for testing. The class-wise support of the test is shown in Fig. \ref{tbl1} The experiments of all models were conducted on Google Colab \cite{Bisong2019}.

\begin{itemize}
    \item Epoch: As the pretrained LMs consisted of hundreds of millions of parameters, we limited training up to 5 epochs, owing to memory limitations \cite{pan2009survey}. 
    \item Batch size: Different batch sizes were used among [16, 32, 64].
    \item Optimiser: we used Adam optimiser with weight decay. \cite{loshchilov2018decoupled}.
    \item Loss weights: We give equal importance to two tasks and add losses as shown in Equation \ref{naive_addition}.

\end{itemize}

%\subsection{Text Preprocessing}
%Preprocessing is one of the important steps in any NLP systems, especially for text classification tasks \cite{UYSAL2014104}. Comments on YouTube are grammatically inconsistent in nature and thus contain emojis, mentions, hashtags, elongated words, expressions, and URLs that makes the process of tokenisation quite strenuous \cite{PLAZADELARCO2021114120}. To overcome these challenges, we follow these steps:
%\begin{itemize}
%    \item The comments are converted to lower case.
%    \item The emojis are replaced by the words that the emoji represents like happy and sad, among
%    other emotions depicted by emojis. As emojis primarily depict the intention of a user,
%    it would be imperative to replace them with their meanings to pick up their cues. As
%    majority of the models are pretrained only on unlabelled text, we feel that it would be necessary.
%    \item URLs and other links are replaced by the word, `URL’.
%    \item Multiple spaces in a sentence and other special characters are removed as they do not contribute significantly in the overall intention of a comment.

%\end{itemize}
\begin{table*}[htbp]
\begin{tabular}{lll}
\hline\noalign{\smallskip}
 
Hyper-parameters & Characteristics  \\
\noalign{\smallskip}\hline\noalign{\smallskip}
Shared layers &  STL: Pooler output from pretrained LM, 1 Linear layer (hidden\_size,128) with a ReLU\\ & activation function, output layer with (128, n\_classes) for the tasks.   \\
& MTL: Same characteristics as STL\\
Task-specific layers &  MTL: 1 linear layer of shape (128, n\_classes) depending on the number of labels\\ & in the datasets.\\
Output layers & STL: 1 Linear layer each for sentiment task ([5] neurons each) and Offensive \\ &task [6], [5], [6] neurons for Kannada, Malayalam, and Tamil, respectively.  \\
& MTL: 2 Linear layers for both tasks ([5,6], [5,5], [5,6] output neurons for\\ & Kannada, Malayalam, and Tamil, respectively)\\
Loss & [CE, KLD, HL, FL] \\ 
Epoch & 5   \\
Batch size & [16, 32, 64]   \\
Optimiser &  AdamW \cite{loshchilov2018decoupled} \\
Dropout  &  0.4 \cite{10.5555/2627435.2670313} \\
Loss weights & [1, 1] (All tasks are treated equally.)\\
\noalign{\smallskip}\hline
\end{tabular}
\caption{Various hyper-parameters used for our experiments}\label{params}
\end{table*}

\subsection{Transfer Learning Fine-Tuning}
We have exhaustively implemented several pretrained language models by fine-tuning them for text classification. % To sum up what has been stated in the previous sections regarding the models, we have fine-tuned a total of eight pretrained models. 
All of the models we use are pretrained on large corpora consisting of unlabelled text. As we are dealing with code-mixed text, it would be interesting to see the performance of the models in Kannada, Malayalam, and Tamil; as all models are pretrained on either mono-lingual or multi-lingual corpora, it is imperative that the fine-tuned models could have a difficulty to classify code-mixed sentences. For the optimiser, we leverage weight decay in Adam optimiser (AdamW), by decoupling weight decay from the gradient update \cite{loshchilov2018decoupled,kingma2014adam}. The primary step is to use the pretrained tokeniser to first cleave the word into tokens. Then, we add the special tokens needed for sentence classification ([CLS] at the first position, and [SEP] at the end of the sentence as shown in Fig. \ref{Fig: BERT_input_representations}). In the figure, tokens T1, T2, ..., Tn represent the cleaved tokens obtained after tokenising.
After special tokens are added, the tokeniser replaces each token with its id from the embedding table which is a component we obtain from the pretrained model.
 
BERT \cite{devlin-etal-2019-bert} was originally pretrained on English texts and was later extended for mbert \cite{pires-etal-2019-multilingual}, which is a language model pretrained on the Wikipedia dumps of the top 104 languages. mBERT consists of 512 input tokens, output being represented as a 768 dimensional vector and 12 attention heads. It is worth noting that we have used both multi-lingual and mono-lingual models (pretrained in English) to analyse the improvements, as we are dealing with code-mixed texts. There are two models of mBERT that are available, for our task, we use the BERT-base, Multi-lingual Cased checkpoint\footnote{\url{https://github.com/google-research/bert/blob/master/multilingual.md}}.
In MTL, we use \textbf{distilbert-base-multilingual-cased} for Tamil while \textbf{bert-base-multilingual-cased} on Malayalam and Kannada datasets, based on their performances on STL on the DravidianCodeMix data set.

%start with why we focus more on weighted avg f1 score than other metrics
%like it explains how good the model is rather than just accuracy
%aim of mtl like why we go for it
%mainly to reduce time and space, soft parameter doesnt reduce space but reduces time
%tell how much space was used during stl as a proof and how much for hard parameter..time for training also if possible
%experiment with a small hope of getting slightly better results in mtl
%make sections for kannada malayalam and tamil again
\section{Results}
\label{Section 7}
This section entails a comprehensive analysis of the capabilities of several pretrained LMs. We have employed the popular metrics in NLP tasks, including precision (P), recall (R), F1-score (F), weighted average, and macro-average. F1-Score is the harmonic average of recall and precision, taking values between 0 and 1. The metrics are computed as follows:\\
\begin{equation}
Precision (c) = \frac{TP}{TP + FP} 
\end{equation}
\begin{equation}
Recall (c) = \frac{TP}{TP + FN} 
\end{equation}
\begin{equation}
F1-score (c) = \frac{2*P(c)*R(c)}{P(c)+R(c)}
\end{equation}
where c is the number of classes:
\begin{itemize}

\item TP – True positive examples are predicted to be positive and are positive;
\item TN – True negative examples are predicted to be negative and are negative;
\item FP – False positive examples are predicted to be positive but are negative; and
\item FN – False negative examples are predicted to be negative but are positive.
\end{itemize}
The weighted average and macro-average are computed as follows:
\begin{equation}
Precision_{M-Avg} = \frac{\sum_{i}^c TP_i}{\sum_{i}^c(TP_i+FP_i)}
\end{equation}  
\begin{equation}
Recall_{M-Avg} = \frac{\sum_{i}^c TP_i}{\sum_{i}^c(TP_i+FN_i)} 
\end{equation}
 
We have evaluated the performance of the models on various metrics such as precision, recall, and F1-score. Accuracy gives more importance to the TP and TN, while disregarding FN and FP. F1-score is the harmonic average of precision and recall. Therefore, this score considers both FP and FN. Due to the persistence of class imbalance in our datasets, we use weighted F1-score as the evaluation metric. The benefits of MTL is multi-fold as these two tasks are related to each other. MTL offers several advantages such as improved data efficiency, reduces overfitting while faster learning by leveraging auxillary representations \cite{DBLP:journals/corr/Ruder17a}. As a result, MTL allows us to have a single shared model in lieu of training independent models per task \cite{dobrescu2020doing}. Hard parameter sharing involves the practice of sharing model weights between multiple tasks, as they are trained jointly to minimise multiple losses. However, soft parameter sharing involves all individual task-specific models to have different weights which would add the distance between the different task-specific models that would have to be optimised \cite{crawshaw2020multi}. We intend to experiment with MTL to achieve a slight improvement in the performance of the model along with reduced time and space constraints in hard parameter sharing.
\begin{table*}[ht]
 
\begin{tabular}{l|rrrr|rrrr}
\hline
\multicolumn{1}{c|}{} & \multicolumn{4}{c|}{Hard Parameter Sharing} & \multicolumn{4}{c}{Soft Parameter Sharing} \\ \hline
Losses & Cross-entropy & Hinge loss & Focal loss & KLD  & CE &  HL &  FL  & KLD\\ \hline
%\hline
\multicolumn{9}{c}{Precision (sentiment analysis)}\\ \hline
%\hline
Positive & 0.726 & 0.448 & 0.453 & 0.426 & 0.450 & 0.620 &0.660 &0.426\\
Negative & 0.453 &0.0 & 0.0&0.0 &0.0 & 0.0 &0.521 &0.0 \\
Mixed feelings &0.0 & 0.0&0.0 & 0.0&0.0 &0.0 & 0.0& 0.0 \\
Neutral & 0.709 & 0.0& 0.0&0.0 & 0.0 &0.595 &0.743 &0.0  \\
Other languages & 0.745 & 0.653& 0.613 & 0.0& 0.616& 0.724&0.758 & 0.0 \\ \hline
Macro-average & 0.527 & 0.220& 0.213&0.085 & 0.213&0.388 &\textbf{0.536} & 0.085\\
Weighted average &\textbf{0.647} & 0.239& 0.238 &0.181 &0.237 &0.515 & 0.638& 0.181\\ \hline
%\hline
\multicolumn{9}{c}{Precision (Offensive language identification)}\\ \hline
%\hline
Not offensive & 0.939 &0.930  &  0.930 &  0.887& 0.933  &0.933   &0.939  &0.887\\
Offensive untargeted &0.0 &0.0 & 0.0& 0.0& 0.0& 0.0 &0.0 &0.0 \\
Offensive targeted individual &0.0 &0.0 &0.0 & 0.0&0.0 & 0.0 &0.0  &0.0  \\
Offensive targeted group &  0.0& 0.0& &0.0 & 0.0&  0.0&0.0  &0.0   \\
Other languages &0.750  &0.635 & 0.585& 0.0&0.714 & 0.780& 0.789&0.0   \\  \hline
%\hline
Macro-average &0.338  &0.313 &0.303 & 0.177& 0.330& {0.343}& \textbf{ 0.345}& 0.177 \\
Weighted average & 0.886 & 0.870 &0.866 & 0.787&0.879 & 0.883&\textbf{0.888} & 0.787 \\  \hline
%\hline
\multicolumn{9}{c}{Recall (sentiment analysis)}\\  \hline
%\hline
Positive &  0.800 & 0.972  &0.965   & 1.000  &0.965   &0.801 &0.871  & 1.000\\
Negative &0.496 & 0.0& 0.0& 0.0&0.0 & 0.0&0.370 &0.0 \\
Mixed feelings &0.0 &0.0 & 0.0& 0.0 &0.0 &0.0 &0.0 &0.0  \\
Neutral & 0.724 & 0.0&0.0 &0.0 &0.0 & 0.670&0.653 & 0.0  \\
Other language & 0.777 &0.681 & 0.777&0.0 &0.734 & 0.755& 0.734& 0.0  \\  \hline
Macro-average & \textbf{ 0.559} &0.331 &0.348 &0.200 &0.340 &0.445 & 0.526& 0.200 \\
Weighted average &\textbf{  0.690} &0.464 & 0.468& 0.426& 0.465& 0.619&0.681 &0.426  \\  \hline
%\hline
\multicolumn{9}{c}{Recall (Offensive language identification)}\\  \hline
%\hline
Not offensive & 0.979 &0.969 &0.961   & 1.000 &0.977   & 0.984  &  0.983 & 1.000\\
Offensive untargeted & 0.0&0.0 &0.0  &0.0 &0.0 & 0.0&0.0 &0.0 \\
Offensive targeted individual &0.0 &0.0 & 0.0 &0.0 & 0.0&0.0 &0.0 & 0.0 \\
Offensive targeted group & 0.0 &0.0 & 0.0 & 0.0& 0.0&0.0 &0.0  &  0.0 \\
Other languages &  0.800&0.678 &0.689 &0.0 &  0.722 &0.711 & 0.789 &  0.0 \\ 
\hline
Macro-average & \textbf{ 0.356} & 0.329& 0.330& 0.200&0.340 & 0.339 &0.354 & 0.200 \\
Weighted average &  0.925& 0.908&0.901&0.887 &0.918 &0.923 &\textbf{ 0.928} & 0.887 \\  \hline
%\hline
\multicolumn{9}{c}{F1-score (sentiment analysis)}\\ \hline
%\hline
Positive & 0.761  &  0.614 & 0.617  &0.597  &0.614   & 0.699  & 0.751 & 0.597\\
Negative &0.473 &0.0 &0.0 &0.0 &0.0 & 0.0&0.433 &0.0 \\
Mixed feelings &0.0 &0.0 &0.0 &0.0 &0.0 &0.0 & 0.0& 0.0 \\
Neutral & 0.716 &0.0 &0.0 &0.0 &0.0 & 0.630 &0.695 & 0.0  \\
Other language & 0.760 &0.667 & 0.685&0.0 & 0.670& 0.740& 0.746&  0.0 \\  \hline
Macro-average & \textbf{ 0.542} & 0.256& 0.260& 0.119& 0.257& 0.414& 0.525& 0.119  \\
Weighted average & \textbf{0.668} &0.310 & 0.313 &0.254 &0.311 &0.561 &0.651 & 0.254 \\  \hline
%\hline
\multicolumn{9}{c}{F1-score (offensive language identification)}\\  \hline
%\hline
Not offensive &  0.959 & 0.949 & 0.945  & 0.940 & 0.955  & 0.958  & 0.960 & 0.940\\
Offensive untargeted &0.0 &0.0 &0.0 & 0.0& 0.0&0.0 &0.00.350 &0.0 \\
Offensive targeted individual &0.0 &0.0 &0.0 & 0.0& 0.0&0.0 &0.0 & 0.0 \\
Offensive targeted group & 0.0 &0.0 &0.0 &0.0 &0.0 & 0.0&0.0 &  0.0 \\
Other languages &0.774  & 0.656 &0.633 &0.0 &0.718 &0.744 &0.789 & 0.0  \\  \hline
%\hline
Macro-average & 0.347 & 0.321& 0.316& 0.188& 0.335&0.340 & \textbf{0.350}& 0.188 \\
Weighted average & 0.905  & 0.888&0.883 &0.834 &0.898 &0.902 &\textbf{0.908} & 0.834 \\ 
\hline
\end{tabular}
\caption{MTL results on the Malayalam data set\\
CE: Cross-entropy loss, HL: Multi-class hinge loss, FL: Focal loss, KLD: Kullback–Leibler divergence}\label{mtl_mal}
\end{table*}

\begin{table*}[ht]
 
\begin{tabular}{l|rrrr|rrrr}
\hline
\multicolumn{1}{c|}{} & \multicolumn{4}{c|}{Hard Parameter Sharing} & \multicolumn{4}{c}{Soft Parameter Sharing} \\ \hline
Losses & Cross-entropy  &  Hinge loss &  Focal loss &  KLD    & CE &   HL  &   FL   &  KLD\\ \hline
\multicolumn{9}{c}{Precision (sentiment analysis)}\\  \hline  
Positive    &  0.721      &  0.703            &  0.729        &  0.560         &  0.710     & 0.686       &  0.717       &  0.676        \\
{Negative}                      & {0.446}    & {0.467}           & {0.449}       & {0.500}        & {0.430}    & {0.444}      & {0.428}      & {0.505}         \\
{Mixed feelings}                & {0.443}     & {0.396}           & {0.389}       & {0.428}        & {0.370}    & {0.419}      & {0.292}      & {0.307}         \\
{Neutral}                 & {0.501}     & {0.523}           & {0.464}       & {0.0}        & {0.540}    & {0.599}      & {0.511}      & {0.540}         \\
{Other language}                & {0.669}     & {0.695}           & {0.667}       & {0.0}        & {0.663}    & {0.721}      & {0.570}      & {0.543}         \\ \hline

{Macro-average}                       & {0.556}     & \textbf{0.557}           & {0.537}       & {0.112}        & {{0.543}}    & {0.583}      & {0.504}      & {0.514}         \\
{Weighted average}                & \textbf{0.619}     & {0.612}           & {0.613}       & {0.313}        & { { 0.610}}    & {0.614}      & {0.596}      & {0.587}         \\ \hline
\multicolumn{9}{c}{ Precision (Offensive language identification)}                                                                                                                                                                                                                                   \\ \hline
{Not offensive}                 & {0.844}     & {0.805}           & {0.843}       & {0.726}        & {0.862}    & {0.821}      & {0.870}      & {0.839}         \\
{Offensive untargeted}          & {0.408}     & {0.4310}           & {0.370}       & {0.0}        & {0.392}    & {0.438}      & {0.395}      & {0.368}         \\
{Offensive targeted individual} & {0.341}     & {0.462}           & {0.335}       & {0.0}        & {0.371}    & {0.418}      & {0.318}      & {0.349}         \\
{Offensive targeted group}      & {0.353}     & {0.373}           & {0.317}       & {0.0}        & {0.410}    & {0.420}      & {0.327}      & {0.323}         \\
{Offensive targeted others}      & {0.0}     & {0.0}           & {0.0}       & {0.0}        & {0.0}    & {0.0}      & {0.0}      & {0.0}         \\
{Other languages}                & {0.823}     & {0.834}           & {0.828}       & {0.0}        & {0.761}    & {0.856}      & {0.850}      & {0.830}         \\ \hline

{Macro-average}                       & {0.461}     & \textbf{0.485}           & {0.449}       & {0.121}        & {0.466}    & {0.492}      & {\textbf{0.460}}      & {0.451}         \\
{Weighted average}                & {0.728}     & {0.713}           & {0.721}       & {0.527}        & \textbf{0.744}    & {0.726}      & {\textbf{0.744}}      & {0.719}         \\ \hline

\multicolumn{9}{ c}{Recall (sentiment analysis)}                                                                                                                                                                                                                                              \\ \hline
{Positive}                      & {0.844}     & {0.867}           & {0.830}       & {1.0}        & {0.857}    & {0.928}      & {0.831}      & {1.0}         \\
{Negative}                      & {0.461}     & {0.433}           & {0.465}       & {0}        & {0.429}    & {0.278}      & {0.342}      & {0}         \\
{Mixed feelings}                & {0.168}     & {0.155}           & {0.155}       & {0}        & {0.168}    & {0.159}      & {0.241}      & {0}         \\
{Unknown state}                       & {0.420}     & {0.372}           & {0.433}       & {0}        & {0.366}    & {0.289}      & {0.356}      & {0}         \\
{Other languages}                & {0.568}     & {0.568}           & {0.587}       & {0}        & {0.601}    & {0.559}      & {0.615}      & {0}         \\ \hline

{Macro-average}                       & {0.492}     & {0.479}           & \textbf{0.494}       & {0.200}        & {0.484}    & {0.443}      & { {0.477}}      & {0.200}         \\
{Weighted average}                & {0.643}     & {0.643}           & {0.637}       & {0.560}        & { {0.639}}    & \textbf{0.645}      & {0.620}      & {0.560}         \\ \hline
\multicolumn{9}{ c}{Recall (Offensive language identification)}                                                                                                                                                                                                                                   \\ \hline
{Not offensive}                 & {0.927}     & {0.959}           & {0.919}       & {1.000}        & {0.909}    & {0.945}      & {0.881}      & {1.000}         \\
{Offensive untargeted}          & {0.443}     & {0.357}           & {0.414}       & {0}        & {0.485}    & {0.375}      & {0.435}      & {0}         \\
{Offensive targeted individual} & {0.196}     & {0.114}           & {0.212}       & {0}        & {0.269}    & {0.177}      & {0.323}      & {0}         \\
{Offensive targeted group}      & {0.200}     & {0.102}           & {0.174}       & {0}        & {0.252}    & {0.190}      & {0.325}      & {0}         \\
{Offensive targeted others}      & {0.0}     & {0.0}           & {0.0}       & {0}        & {0.0}    & {0.945}      & {0.0}      & {0.0}         \\
{Other language}                & {0.730}     & {0.708}           & {0.730}       & {0}        & {0.787}    & {0.736}      & {0.730}      & {0}         \\ \hline

{Macro-average}                       & {0.416}     & {0.373}           & {0.408}       & {0.167}        & \textbf{0.450}    & {0.404}      & {{0.449}}      & {0.167}         \\
{Weighted average}                & {0.766}     & \textbf{0.769}           & {0.757}       & {0.726}        & {0.767}    & {0.772}      & { {0.750}}      & {0.726}         \\ \hline

\multicolumn{9}{ c}{F1-score (sentiment analysis)}                                                                                                                                                                                                                                              \\ \hline
{Positive}                      & {0.778}     & {0.777}           & {0.777}       & {0.718}        & {0.777}    & {0.778}      & {0.770}      & {0.718}         \\
{Negative}                      & {0.454}     & {0.449}           & {0.457}       & {0.0}        & {0.430}    & {0.357}      & {0.380}      & {0}         \\
{Mixed feelings}                & {0.243}     & {0.223}           & {0.222}       & {0.0}        & {0.231}    & {0.232}      & {0.264}      & {0}         \\
{Neutral}                       & {0.457}     & {0.435}           & {0.450}       & {0.0}        & {0.436}    & {0.390}      & {0.420}      & {0}         \\
{Other language}                & {0.614}     & {0.625}           & {0.616}       & {0.0}        & {0.631}    & {0.630}      & {0.591}      & {0}         \\ \hline

{Macro-average}                       & \textbf{0.509}     & {0.502}           & {0.504}       & {0.144}        & {0.501}    & {0.477}      & {\textbf{0.485}}      & {0.144}         \\
{Weighted average}                & {0.621}     & {0.614}           & \textbf{0.625}       & {0.402}        & {0.614}    & {0.598}      & {\ {0.603}}      & {0.402}         \\ \hline
\multicolumn{9}{ c}{F1-score (offensive language identification)}                                                                                                                                                                                                                                   \\ \hline
{Not offensive}                 & {0.883}     & {0.875}           & {0.879}       & {0.841}        & {0.885}    & {0.879}      & {0.876}      & {0.841}         \\
{Offensive untargeted}          & {0.425}     & {0.393}           & {0.390}       & {0}        & {0.434}    & {0.404}      & {0.414}      & {0}         \\
{Offensive targeted individual} & {0.249}     & {0.183}           & {0.260}       & {0}        & {0.312}    & {0.249}      & {0.320}      & {0}         \\
{Offensive targeted group}      & {0.255}     & {0.160}           & {0.225}       & {0}        & {0.312}    & {0.262}      & {0.326}      & {0}         \\
{Offensive targeted others}      & {0.0}     & {0.0}           & {0.000}       & {0}        & {0.0}    & {0.0}      & {0}      & {0}         \\
{Other language}                & {0.774}     & {0.766}           & {0.776}       & {0}        & {0.773}    & {0.792}      & {0.785}      & {0}         \\ \hline

{Macro-average}                       & {0.431}     & {0.396}           & {0.422}       & {0.140}        & \textbf{0.453}    & {0.431}      & {\textbf{0.453}}      & {0.140}         \\
{Weighted average}                & {0.742}     & {0.729}           & {0.745}       & {0.611}        & \textbf{0.753}    & {0.739}      &   {0.746}       & {0.611}         \\ \hline
                                              
\end{tabular}
\caption{MTL on the Tamil data set}\label{mtl_tam}
\end{table*}

Apart from the pretrained LMs shown in Table\,\ref{stl_kan}, \ref{stl_mal}, and \ref{stl_tam}, we have tried other domain specific LMs. IndicBERT \cite{kakwani-etal-2020-indicnlpsuite} is a fastText-based word embeddings and ALBERT-based LM trained on 11 Indian languages along with English and is a multi-lingual model similar to mBERT \cite{pires-etal-2019-multilingual}. IndicBERT was pretrained on IndicCorp \cite{kakwani-etal-2020-indicnlpsuite}, the sentences of which were trained using a sentence piece tokeniser \cite{kudo-richardson-2018-sentencepiece}. However, when IndicBERT was fine-tuned for sentiment SA and OLI, it was found that the model performed very poorly, despite being pretrained on 12 languages, inclusive of Kannada, Malayalam, Tamil, and English. We believe that one of the main reasons for its poor performance, despite being pretrained on a large corpus, has to do with the architecture of the model, which was also encountered by \cite{adeephope}. Even though IndicBERT was pretrained in a multi-lingual setting, it followed the architecture of ALBERT using the standard MLM objective. We believe that this is due to the cross-layer parameter sharing that hinders its performance in a multi-lingual setting (when its pretrained on more than one language). Consequentially, ALBERT tends to focus on the overall spatial complexity and training time \cite{DBLP:journals/corr/abs-1909-11942}.

We have also experimented with multi-lingual Representations for Indian languages (MuRIL), a pretrained LM that was pretrained on Indian languages as similar to IndicBERT but differs in terms of pretraining strategy and the corpus used \cite{khanuja2021muril}. It follows the architecture of BERT-base encoder model; however, it is trained on two objectives, MLM, and transliterated language modeling (TLM). TLM leverages parallel data unlike the former training objective. However, in spite of having a TLM objective, the model performed worse than BERT base. which was the base encoder model for MuRIL. Hence, we have not tabulated the classification report of both MuRIL and IndicBERT.
 We will be analysing the performance of models on different datasets separately.
% \section{Results}
\subsection{Kannada}
 
Table\,\ref{stl_kan} and Table\,\ref{mtl_kan} present the classification report of the models on Kannada data set. We opted to train the STL with CE loss, which is a popular choice among loss functions for classification instances \cite{de2005tutorial}. We observe that mBERT (multi-lingual-bert) achieves the highest weighted F1-score among other models as observed in Table\,\ref{stl_kan}. 
\begin{figure}[htbp]
\center\includegraphics[width=8cm,height=6cm]{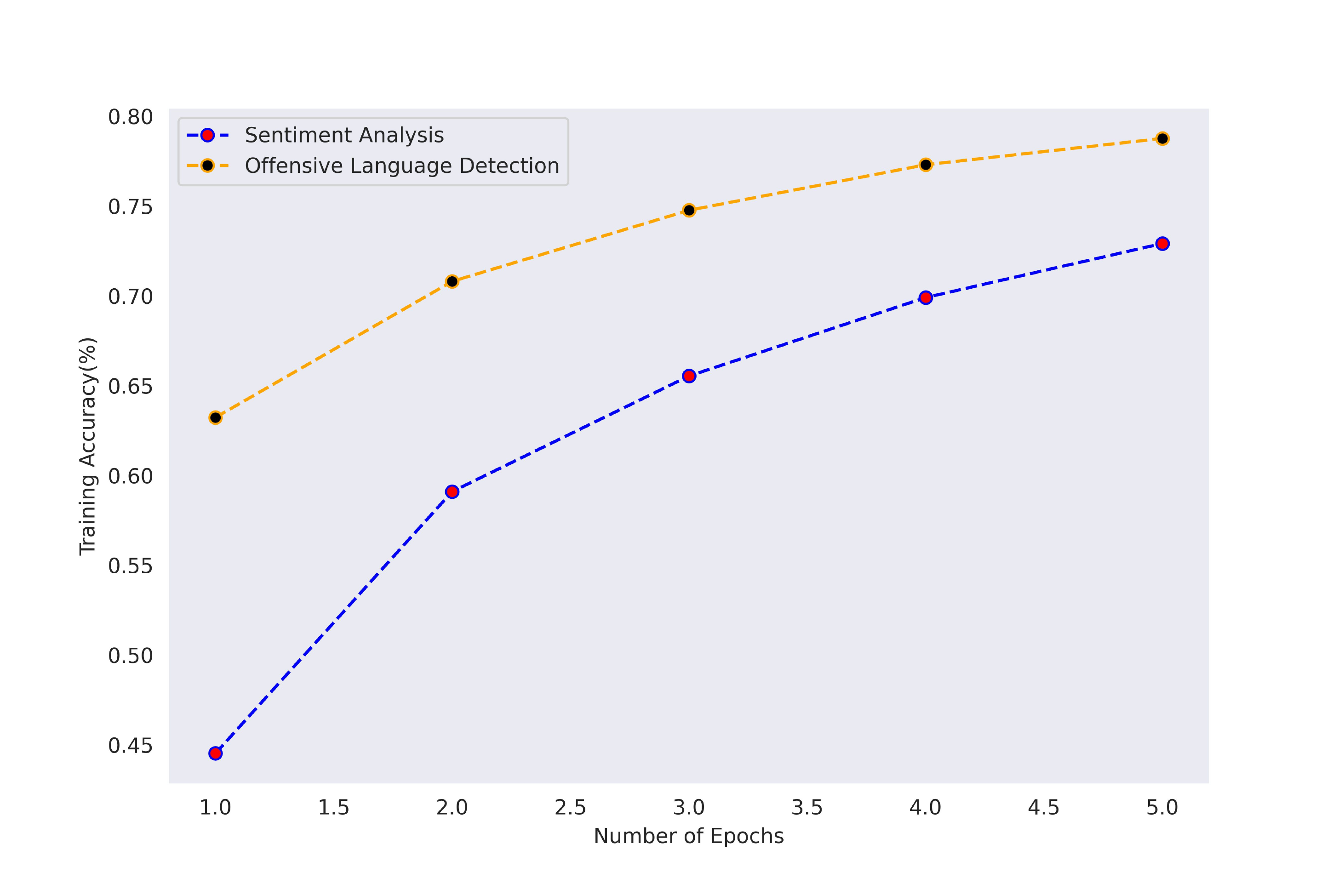}
\center\caption{Train Accuracy during MTL}
\label{kan_acc}
\end{figure}
mBERT achieved 0.591 for SA and 0.686 for OLI. In spite of achieving the highest weighted average of F1-scores of the classes, we can observe that the some classes have performed very poorly with 0.0 as their F1-score. In OLI, we can see that most of the models perform poorly on 3 classes: \textit{Offensive Untargeted}, \textit{Offensive Targeted Group}, and \textit{Offensive Targeted Others}. One of the main reasons could be due to the less support of these classes in the test set. \textit{Offensive Untargeted} has a support of 27, while \textit{offensive Targeted Group} has 44, and \emph{Offensive Targeted others} has a mere 14 out of the total test set having a support of 728. As the ratio of the majority class to the minority class is very severe (407:14), which is essentially the reason why the weighted F1-scores of \textit{Not Offensive} and \textit{Offensive Targeted Individual} are much higher in comparison to the low-resourced classes. Among mono-lingual LMs, characterBERT performs at par with mBERT. It is interesting to note that we use \emph{general-character-bert}\footnote{\url{https://github.com/helboukkouri/character-bert\#pretrained-models}}, a LM that was pretrained only on English Wikipedia and OpenWebText \cite{liu-curran-2006-web}. We believe the approach employed in characterBERT, of attending to the characters to attain a single word embedding as opposed to the process of tokenisation in BERT, as illustrated in Fig. \ref{Fig:characterBERT}. This is one of the reasons why characterBERT outperforms other mono-lingual models such as RoBERTa, ALBERT, and XLNet. Despite the superiority of other models that are pretrained on more data and better strategies.

\begin{figure}
     \centering
     \begin{subfigure}[!ht]{0.5\textwidth}
         \centering
         \includegraphics[width=\textwidth]{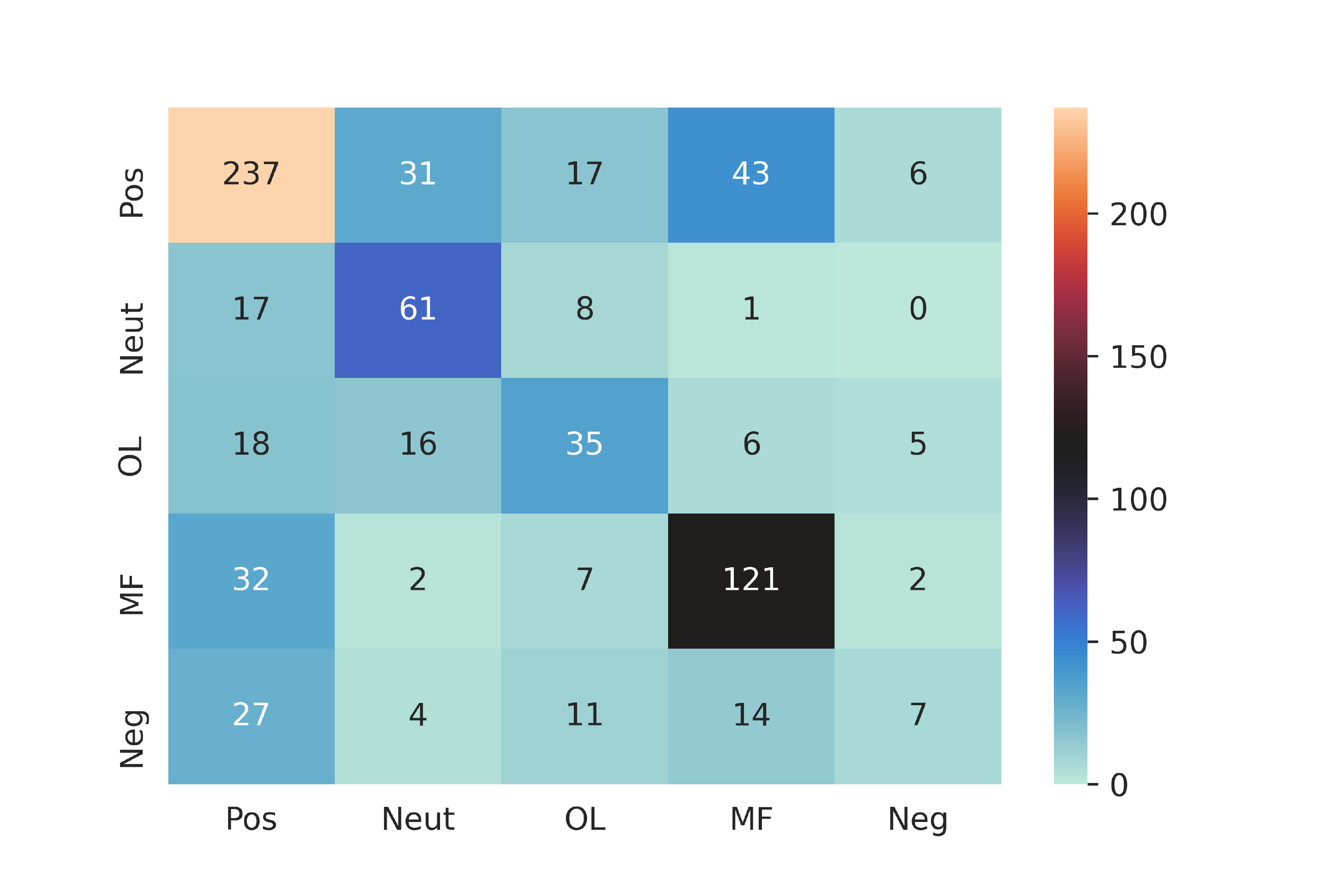}
         \caption{Pos: Positive, Neut: Neutral, OL: Other languages, MF: Mixed feelings, Neg: Negative}
         \label{kan_sent}
     \end{subfigure}
     \hfill
     \begin{subfigure}[!ht]{0.5\textwidth}
         \centering
         \includegraphics[width=\textwidth]{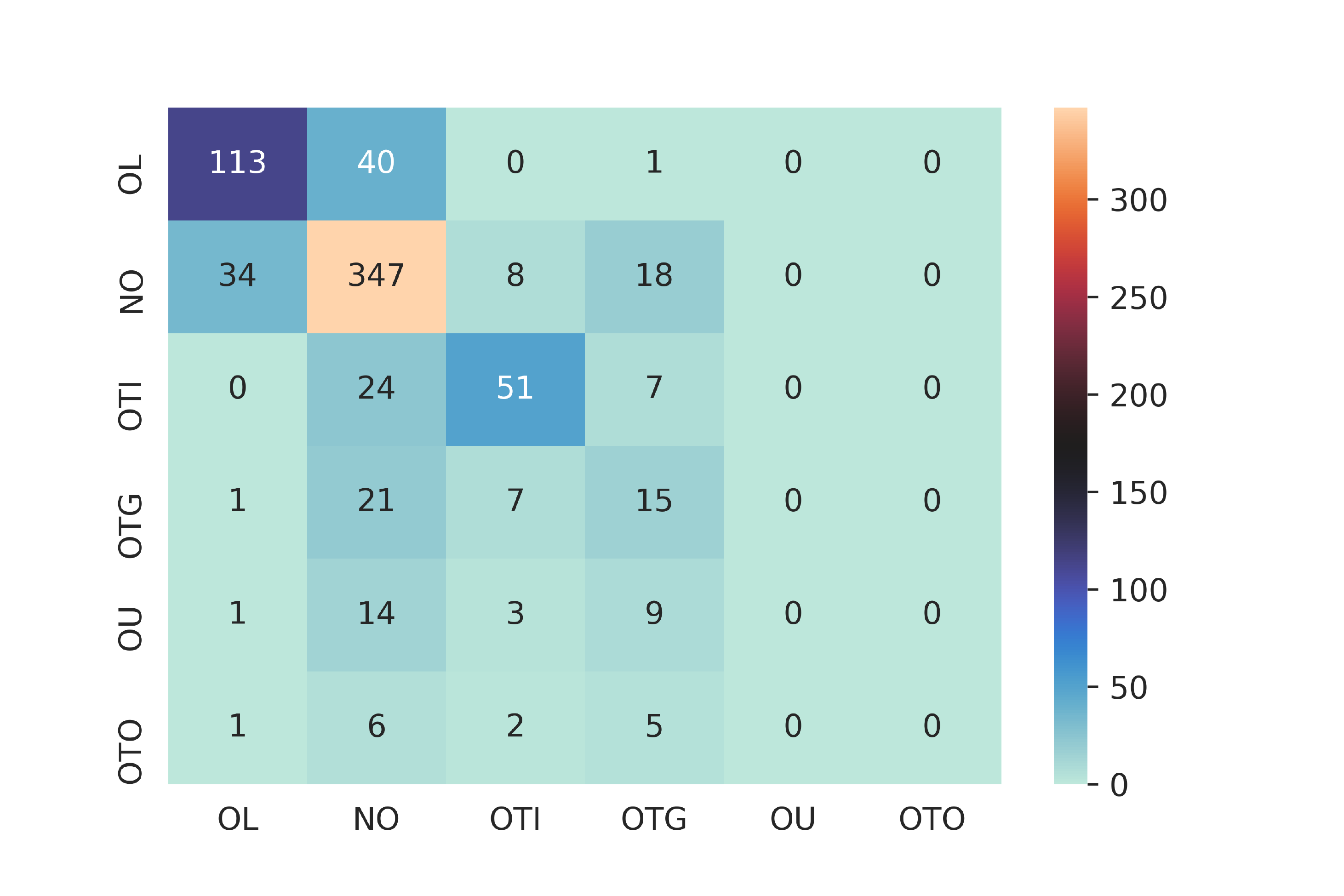}
         \caption{NO: Not offensive, OL: Other language, OU: Offensive untargeted, OTG: Offensive targeted group, OTI: Offensive targeted individual, OTO: Offensive targeted others}
         \label{kan_off}
     \end{subfigure}
    \caption{Heatmap of confusion matrix for the best performing model on the Kannada data set}
    \label{kan_cm}
\end{figure}

To our surprise, we observe that XLM-RoBERTa, a multi-lingual LM pretrained on the top 100 languages, performed the worst among the models. One of the main reasons for the low performance of XLM-RoBERTa (–20\%) could be on the account of low support of the test set. However, due to the nature of the language, it would be hard to extract more data. Another reason for the poor performance is based on the architecture of RoBERTa models, upon which the XLM-R model is based. Unlike the conventional wordpiece \cite{6289079} and unigram tokeniser \cite{kudo-2018-subword}, RoBERTa uses a byte-level BPE tokeniser \cite{sennrich-etal-2016-neural} for its tokenisation process. However, BPE tends to have a poor morphological alignment with the original text \cite{jain2020indictransformers}. As Dravidian languages are morphologically rich \cite{10.1145/3407912}, this approach results in poor performance on SA and OLI. DistilBERT scores more than the other multi-lingual models such as XLM and XLM-R, in spite of having very few parameters in comparison to large LMs such as XLM-R base (66M vs 270M). However, among all models, mBERT had the best overall scores in both tasks. Hence, we used BERT to perform the MTL models by training them for hard parameter sharing and soft parameter sharing.

We have fine-tuned BERT for these loss functions as it outperformed other pretrained LMs for CEs. We have employed four loss functions. The highest weighted F1-score for SA was 0.602, which was achieved when mBERT was trained in soft parameter sharing strategy with CE as the loss function. There is an increase of 0.101 in contrast to STL. However, the output of the second task scored 0.672, which is lower than the STL score (-0.140 F1-score). As we take the naive sum of the losses when training in a multi-task setting, the losses of one task suppresses the other task, thus, only one task majorly benefit from the approach \cite{Maninis2019AttentiveSO}. However, it is worth noting that the approach did achieve a competitive weighted F1-score, if not greater than the former. The highest F1-score achieved for OLI was 0.700, which was achieved by training mBERT in hard parameter sharing strategy, with CE as its loss function. However, the same issue of supression of performance on the other task was observed as it scored 0.590. It can be observed that only two of the classes scored 0.0 in MTL in contrast to three in STL. When both models were trained with KLD as its loss function, the performance was abysmal, and it was only able to classify a single class in the respective tasks (positive and not offensive) among all. When trained on hinge loss and focal loss, we observe that they attend to the class imbalance. Hence, we observe that MTL frameworks tend to perform slightly better than when treated as individual tasks.

Fig. \ref{kan_acc} represents the training accuracy of SA and OLI in an MTL scenario. It can be observed from the graph that there is a bigger jump for SA after the first epoch in contrast to OLI, where there is a steady increase in the accuracy. Fig. \ref{kan_sent} and Fig. \ref{kan_off} exhibit the confusion matrices of the best performing model for the respective tasks. For SA, we observe that a lot of labels are being misclassified. The trend of misclassification can also be observed in OLI, where most of the samples are being misclassified into four classes (instead of 6), with the absence of samples being classified into OTO and OU. The samples of OTO and OU are being misclassified into other classes, which is mainly due to the low support of these classes in the test set.
 \begin{figure}[htbp]
\center\includegraphics[width=8cm,height=6cm]{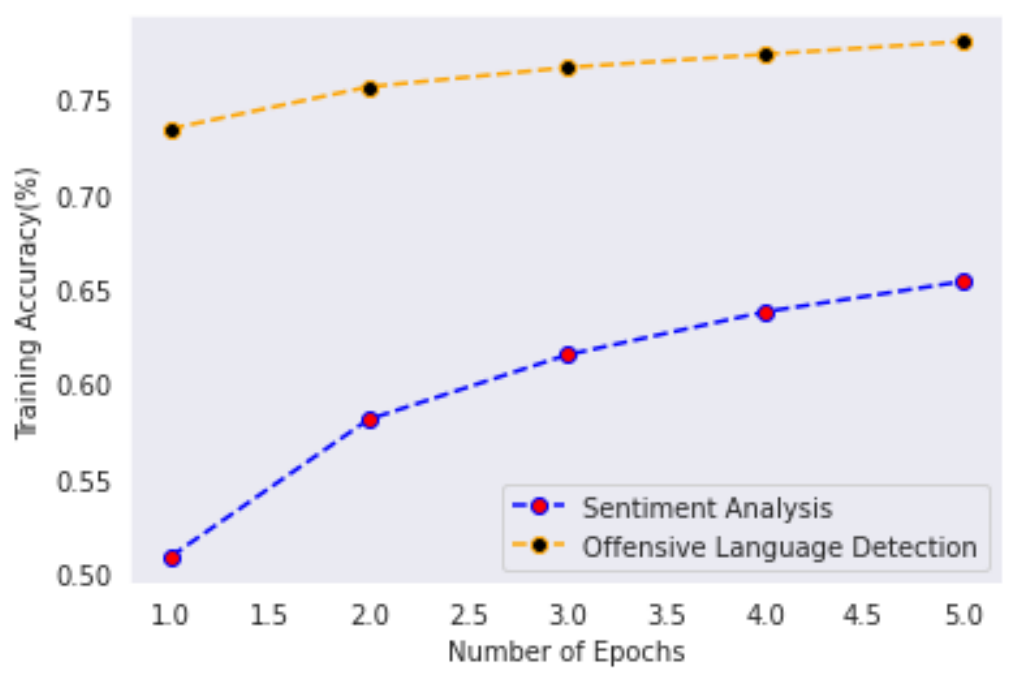}
\center\caption{Train accuracy during MTL of DistilmBERT on the Tamil data set}
\label{tam_acc}
\end{figure}
\subsection{Tamil}
Table\,\ref{tbl1} gives an insight into the support of the classes on the test set. We observe that the support of Offensive Targeted Others is quite low in contrast to not offensive (52:3148). From Table\,\ref{stl_tam}, DistilmBERT is the best performing model among all models. In spite of being a smaller, distilled form of BERT, it performs better than the parent model. Even though it was suggested that DistilBERT retains 97\% of BERT's accuracy, it has performed better than BERT previously \cite{maslej2020comparison}, mainly due to the custom triple loss function and fewer parameters
\cite{DBLP:journals/corr/abs-1909-11942}. The weighted average of F1-score for SA is 0.614 and 0.743 for OLI, all trained with CE. Even though BERT achieves competitive performance, three out of six classes have a weighted average F1-score of 0.0, due to misclassification. Hence, we train DistilmBERT on MTL. We observe that XLM has also performed better than BERT and XLM-RoBERTa.
\begin{figure}
     \centering
     \begin{subfigure}[htbp]{0.4\textwidth}
         \centering
         \includegraphics[width=\textwidth]{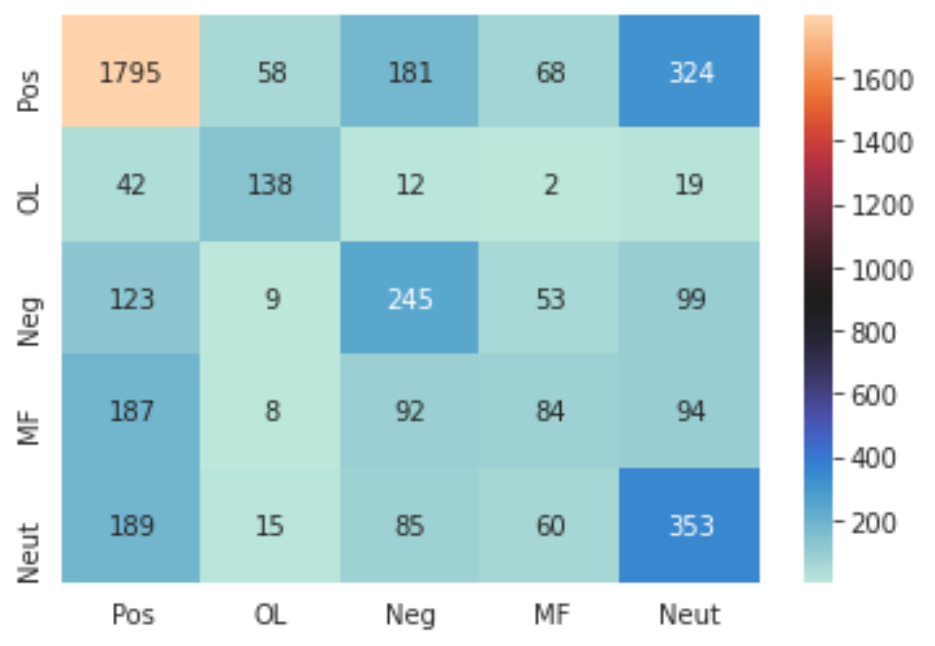}
         \caption{Pos: Positive, Neut: Neutral, OL: Other languages, MF: Mixed feelings, Neg: Negative}
         \label{tam_sent}
     \end{subfigure}
     \hfill
     \begin{subfigure}[htbp]{0.4\textwidth}
         \centering
         \includegraphics[width=\textwidth]{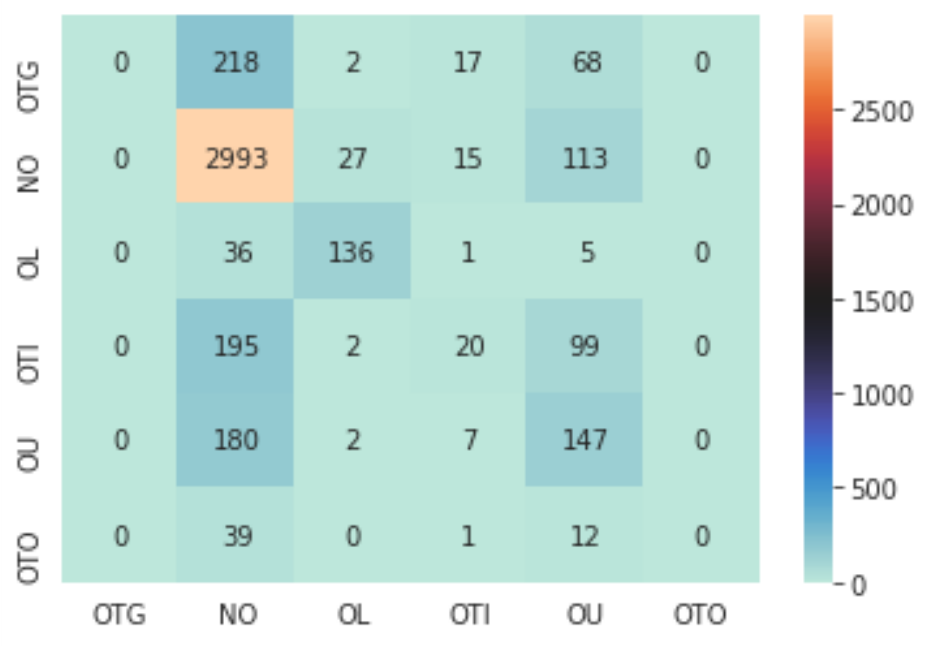}
         \caption{NO: Not offensive, OL: Other languages, OU: Offensive untargeted, OTG: Offensive targeted group, OTI: Offensive targeted individual}
         \label{tam_off}
     \end{subfigure}
      
        \caption{Heatmap of confusion matrix for the best performing model on the Tamil data set}
        \label{fig:tam_cm}
\end{figure}

 %tamil
%similar intro about support but decent support for all classes
%why distilbert over other models? better results
%highest value for task1: hard parameter: 0.625 focal loss
%highest value for task2: soft parameter: 0.753 CE
%hard parameter focal is also nearby 0.745
%soft parameter CE is also nearby 0.614
%so both of these models are goof(if u have any other reason why one of them is better choose that and write reason..i couldnt find any)
%highest value in STL distilbert task1 and task 2: 0.614 and 0.743
%almost similar values in results. only gain is time and space 
%no loss and no gain in results
It can be observed from Table\,\ref{mtl_tam} that DistilmBERT, when trained using hard parameter sharing strategy on CE loss yields the best results for OLI with a weighted F1-score of 0.753 (+1.0\%), and 0.614 on SA (+/-0.0\%). However, the best score for OLI was achieved by training DistilmBERT on soft parameter sharing strategy with Focal Loss. It achieved a score of 0.625 (+1.1\%) on SA and 0.745 (+0.2\%) on OLI. It is to be noted that training on both CE and FL achieves better results than the best scores set forth by the STL model. The hinge loss function helps achieve better results for precision and recall. During training, the accuracy curve of SA is increasing at a greater rate in contrast to OLI, as displayed in Fig. \ref{tam_acc}. Fig. \ref{fig:tam_cm} displays the class-wise predictions by DistilmBERT trained on hard parameter sharing with CE. Most of the samples of NO class are predicted correctly, with the majority of the rest being mislabelled as OU. Due to low support in the training and test set, no samples of OTO have been predicted correctly. Most of the classes are incorrectly predicted as OU, which is primarily due to the close interconnection among the classes. In SA, it can be observed that a significant number of samples of \textit{Positive} and \textit{Negative} classes have been predicted as \textit{Neutral}. This is also observed in \textit{Mixed Feeling} (MF), where most of the samples have been misclassified as \textit{Positive} and \textit{Negative}. The misclassification is mainly due to the nature of the class, as \textit{Mixed Feelings (MF)} can be either of the two classes.

\subsection{Malayalam}
 
\begin{figure}[htbp]
\center\includegraphics[width=8cm,height=6cm]{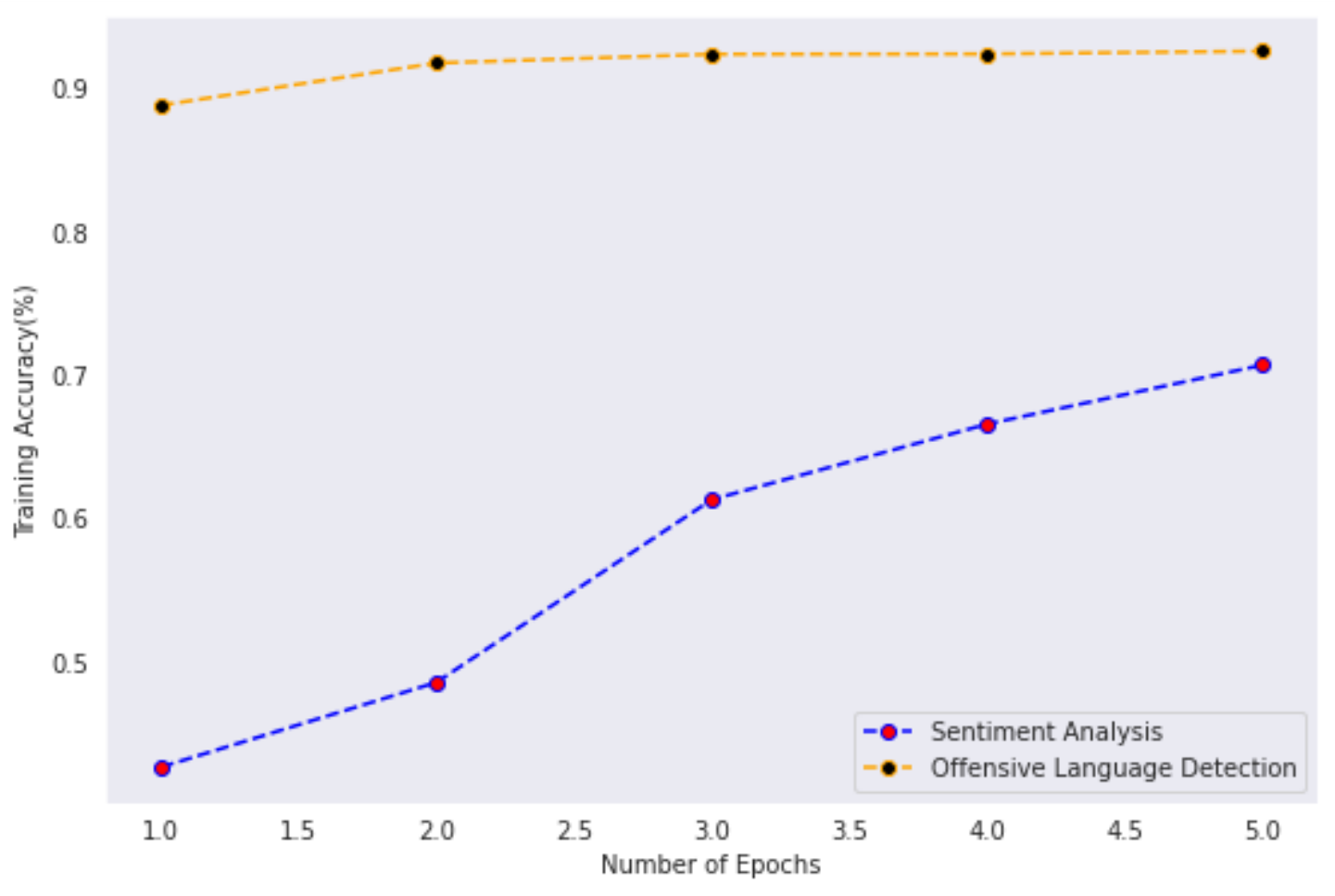}
\center\caption{Train Accuracy during MTL of mBERT on the Malayalam data set}
\label{Fig:mal_acc}
\end{figure}
 
Table\,\ref{stl_mal} and \ref{mtl_mal} illustrate the classification report of the models on STL and MTL, respectively. It can be observed that among STL, fine-tuning mBERT with CE gave a weighted F1-score of 0.635 for SA and 0.902 for OLI. DistilBERT outperforms BERT on OLI, however, it scores poorly on SA. DistilBERT reduces the occurrence of misclassification, as it is able to classify four out of the five classes separately, unlike mBERT, which correctly classifies two out of five classes in OLI. Even though XLM has higher class-wise F1-scores for both tasks, when trained on an MTL setting, it performed very poorly. This could be due to the pretraining strategy of XLM, as it was only pretrained on Wikipedia, which does not improve the performance on low-resource languages. Added to that, acquiring parallel data for TLM during pretraining could be very challenging \cite{gregoire-langlais-2018-extracting}.
Several classes have scored 0.0 for class-wise weighted F1-scores due to less support on the test set (Class Imbalance). Since mBERT is the strongest model among all, we experiment it for MTL. 
 
\begin{figure}
     \centering
     \begin{subfigure}[htbp]{0.4\textwidth}
         \centering
         \includegraphics[width=\textwidth]{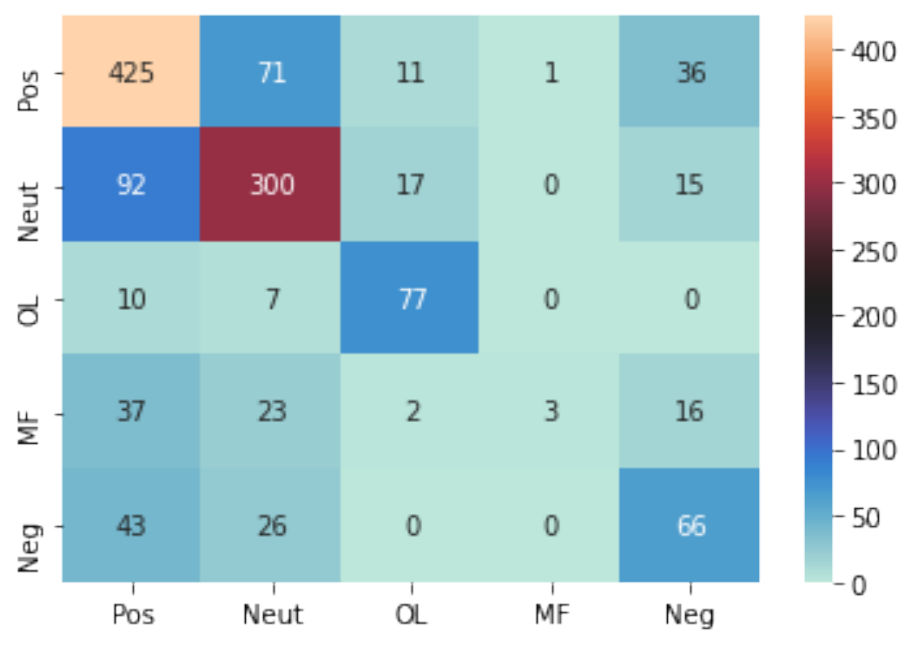}
         \caption{Pos: Positive, Neut: Neutral, OL: Other languages, MF: Mixed feelings, Neg: Negative}
         \label{mal_sent}
     \end{subfigure}
     \hfill
     \begin{subfigure}[htbp]{0.4\textwidth}
         \centering
         \includegraphics[width=\textwidth]{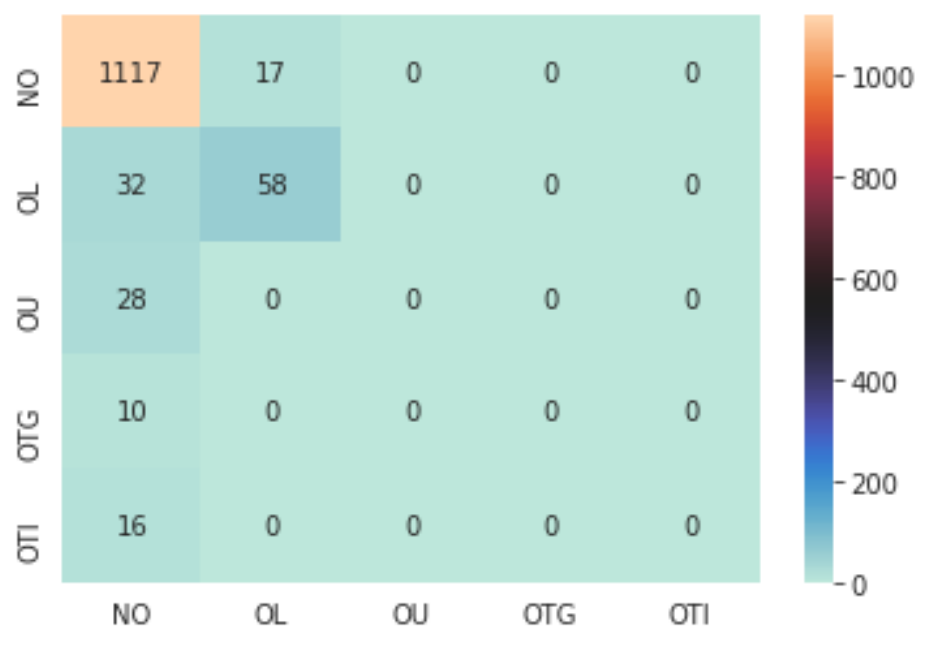}
         \caption{NO: Not offensive, OL: Other languages, OU: Offensive untargeted, OTG: Offensive targeted group, OTI: Offensive tTargeted individual}
         \label{mal_off}
     \end{subfigure}
      
        \caption{Heatmap of confusion matrix for the best performing model on the Malayalam data set}
        \label{fig:mal_cm}
\end{figure}

We observe that mBERT fine-tuned on hard parameter sharing strategy with CE as its loss, achieved the highest score of 0.668 (+3.3\% from STL) in SA, while its secondary task scored 0.905 (+0.03\%). Despite not being the best weighted F1-score, it outperformed the STL result of mBERT. We observe that when mBERT is trained using soft parameter sharing with focal loss being its loss function, it attains the highest weighted F1-score for OLI (+0.6\%), while scoring 0.651 (+1.6\%) on SA. We also observe that KLD's performance is appalling and could be due to the perseverance of the loss function. KLD is likely to treat the following multi-class classification problem as a regression problem, hence predicting a single class that has the highest support. However, it has been previously used for classification problems \cite{nakov-etal-2016-semeval}, but in our case, the loss performs poorer than the performance of the algorithms that serve as the baseline \cite{chakravarthi-etal-2020-sentiment}. It is to be noted that the increase in performance is more for SA in contrast to OLI. The time required to train, along with the memory constraints, is one of the reasons why we opt for MTL.

 Fig. \ref{Fig:mal_acc} represents the training accuracy of both tasks in the best performing model. After the first epoch, we see that there is a steady increase in both of the tasks. However, after the third epoch, we do not see any improvement in the performance of both of the tasks. Fig. \ref{fig:mal_cm} reports the confusion matrices of the best performing models (in MTL). For SA, we observe that most of the labels are misclassified for the samples of mixed feelings, which is mainly due to the low support in the training and test set as observed in Fig. \ref{mal_sent}. For OLI, we observe that the model tends to classify all of the samples in the test set into two classes: NO and OL. As stated previously, low support of the other classes during training is why the model misclassifies its labels.

\section{Conclusion}
\label{Section 8}
Despite of the rising popularity of social media, the lack of code-mixed data in Dravidian languages has motivated us to develop MTL frameworks for SA and OLI in three code-mixed Dravidian corpora, namely, Kannada, Malayalam, and Tamil. The proposed approach of fine-tuning multi-lingual BERT to a hard parameter sharing with cross-entropy loss yields the best performance for both of the tasks in Kannada and Malayalam, achieving competitive scores in contrast to the performance when its counterparts are treated as separate tasks. For Tamil, our approach of fine-tuning multi-lingual distilBERT in soft parameter sharing with cross-entropy loss scores better than the other models, mainly due to its triple loss function employed during pretraining. The performance of these models highlights the advantages of using an MTL model to attend to two tasks at a time, mainly reducing the time required to train the models while additionally reducing the space complexities required to train them separately. For future work, we intend to use uncertainty weighting to calculate the impact of one loss function on the other during MTL.

\begin{acknowledgements}
The author Bharathi Raja Chakravarthi was supported in part by a research grant from Science Foundation Ireland (SFI) under Grant Number SFI/12/RC/2289$\_$P2 (Insight$\_$2), co-funded by the European Regional Development Fund and Irish Research Council grant IRCLA/2017/129 (CARDAMOM-Comparative Deep Models of Language for Minority and Historical Languages).\\
\end{acknowledgements}
\section*{Funding}
This  research  has  not  been  funded  by  any  company  or  organization
\section*{Compliance with Ethical Standards}
\textbf{Conflict of interest:} The authors declare that they have no conflict of interest.\\
\\
\textbf{Availability of data and material:} The datasets used in this paper are obtained from \url{https://github.com/bharathichezhiyan/DravidianCodeMix-Dataset}.\\
\\
\textbf{Code availability:} The data and approaches discussed in this paper are available at \url{https://github.com/SiddhanthHegde/Dravidian-MTL-Benchmarking}.\\
\\
\textbf{Ethical Approval:}  This article does not contain any studies with human participants or animals performed by any of the authors.\\
 
% BibTeX users please use one of
%\bibliographystyle{spbasic}      % basic style, author-year citations
\bibliographystyle{spmpsci}      % mathematics and physical sciences
\bibliography{refs}   % name your BibTeX data base

% Non-BibTeX users please use
 
% \section{Abbreviations}
% The following abbreviations are used in the manuscript:
% \begin{table}[htbp] 
%     \begin{tabular}{ll}
%       mBERT & Multilingual BERT\\
%       SA  & Sentiment Analysis \\
%       OLD  & Offensive Language Detection \\
%       STL &Single-task Learning \\ 
%       MTL &Multi-task Learning \\
%       CE &Cross-Entropy \\
%       LM & Language Models \\
%       HPS & Hard Parameter Sharing\\
%       SPS &Soft Parameter Sharing \\
%     \end{tabular} 
%     \label{abbreviation}
% \end{table}
 
\end{document}